\documentclass[pdflatex,sn-basic]{sn-jnl}

\newcommand{\mycomment}[1]{}

\usepackage{graphicx}%
\usepackage{multirow}%
\usepackage{amsmath,amssymb,amsfonts}%
\usepackage{amsthm}%
\usepackage{mathrsfs}%
\usepackage[title]{appendix}%
\usepackage{xcolor}%
\usepackage{textcomp}%
\usepackage{manyfoot}%
\usepackage{booktabs}%
\usepackage{algorithm}%
\usepackage{algorithmicx}%
\usepackage{algpseudocode}%
\usepackage{listings}%
\usepackage{float}%
\usepackage{placeins}

\usepackage{subcaption}

\usepackage{doi}
\usepackage{csquotes}

\usepackage{csquotes} 

\usepackage[table]{xcolor}
\usepackage{colortbl}

\usepackage{bm}

\usepackage[normalem]{ulem}

\mycomment{
\theoremstyle{thmstyleone}%
\theoremstyle{thmstyletwo}%

\theoremstyle{thmstylethree}%

}

\begin{document}

\title[Spatio-Temporal Dynamics of Obesity in Italian Regions]{Modeling Spatio-Temporal Dynamics of Obesity in Italian Regions Via Bayesian Beta Regression}


\author[1]{\fnm{Luciano} \sur{Rota}}\email{l.rota43@campus.unimib.it}

\author[2]{\fnm{Raffaele} \sur{Argiento}}\email{raffaele.argiento@unibg.it}

\author*[2]{\fnm{Michela} \sur{Cameletti}}\email{michela.cameletti@unibg.it}

\affil[1]{\orgdiv{Dipartimento di Economia, Metodi Quantitativi e Strategie di Impresa}, \orgname{Università di Milano-Bicocca}, \orgaddress{\street{Piazza dell'Ateneo Nuovo, 1}, \city{Milano}, \postcode{20126}, \state{MI}, \country{Italy}}}

\affil[2]{\orgdiv{Dipartimento di Scienze Economiche}, \orgname{Università degli studi di Bergamo}, \orgaddress{\street{Via dei Caniana}, \city{Bergamo}, \postcode{24127}, \state{BG}, \country{Italy}}}

\abstract{In this paper, we investigate the spatio-temporal dynamics of obesity across Italian regions from 2010 to 2022, introducing a novel way to analyse aggregated regional obesity data. In particular, we implement a Bayesian hierarchical Beta regression model to analyse regional obesity rates, integrating spatial and temporal random effects, alongside gender and several exogenous predictors. The model leverages the Stochastic Search Variable Selection technique to identify significant predictors supported by the data.
The analysis reveals both gender and regional heterogeneity in obesity rates over the study period. Structured spatial and temporal random effects, along with gender, emerge as the primary determinants of obesity prevalence across Italian regions, while the role of exogenous covariates is found to be minimal at the regional level, once structured spatial and temporal effects and gender are accounted for.
While socioeconomic and lifestyle factors remain fundamental at a micro-level, the findings suggest that the integration of spatial and temporal structures is crucial for capturing macro-level obesity variations.}




\maketitle

\section{Introduction}\label{sec:intro}

Obesity is recognized as a major public health problem in Europe, where it has reached alarming levels, posing substantial health, social, and economic burdens on European societies. In 2014, one in six adults were considered obese according to the European Health Interview Survey (2016)\footnote{Eurostat news release, 203/2016, 20 October 2016, \url{https://ec.europa.eu/eurostat/documents/2995521/7700898/3-20102016-BP-EN.pdf/c26b037b-d5f3-4c05-89c1-00bf0b98d646}}. Moreover, projections suggest an increasing average obesity trend \citep{janssen2020}. 

Even if the problem is common among countries, such a phenomenon is not homogeneous: overweight and obesity vary substantially between and within countries. 
For instance, compared to other European countries, Italy shows relatively lower obesity rates for women \citep{oecd2022health}. 
According to the Eurostat data \citep{Eurostat2022}, the second lowest proportion of obese people in 2019 was recorded in Italy (slightly above 11\%), when nearby countries like Spain, France, Germany, and Austria are all above 14 \%, and the European mean is around 17\%. Despite this, the situation is still problematic and needs appropriate attention and intervention.

Moreover, national averages hide substantial heterogeneity within the countries, both in space and by gender, through time. Focusing on Italy, between northern and southern regions, there is a notable and time consistent difference in overweight and obesity rates. For example, taking the best and worst regions in the period considered by this study (2010-2022), our data show that the female (male) obesity rate is 7.50\% (9.40\%) in Trentino-Alto Adige and 12.7\% (14.4\%) in Molise. While considering overweight for the same regions, the percentages are 23.5\% (40.3\%) for Trentino-Alto Adige and 31\% (46\%) for Molise.
It is well-known that across Italian regions, there are strong structural differences in general culture, social norms, and economic development (for a general discussion, see, for instance, \citealp{deblasionuzzo2009}). This diversity among Italian regions may influence obesity and related phenomena \citep{gallus2013}. For this reason, proposing an analysis that takes into account heterogeneity is essential \citep{brunello2014}. Flexible models, such as varying coefficient models, have also been proposed in \cite{assaf2015} to explore spatially and temporally varying effects on public health outcomes, including obesity, using large-scale survey data. Their findings show that this framework can capture dynamic changes over space and time in the effects of key risk factors, providing insight into the evolution of obesity across population subgroups. Their focus, however, differs from ours: while they concentrate on time and subgroup-specific variation in covariate effects, we are mainly interested in studying and interpreting the spatio-temporal pattern of obesity rates after accounting for covariates. 
Along with space and time, gender aspects should also be considered, given potential physical and genetic differences \citep{vari2016} as well as cultural ones \citep{barrabarone2021}.

In this paper, we analyse longitudinal areal data consisting of yearly regional obesity rates from 2010 to 2022, which were assembled through a systematic process of data collection, validation, and harmonization across multiple institutional sources.
To model these rates, expressed as proportions bounded between 0 and 1, we employ a Beta regression model within a Bayesian hierarchical framework, using a mean–precision parameterization. This approach provides a flexible structure for modelling rate data, enabling the inclusion of covariates as well as unobserved spatial and temporal heterogeneity through structured random effects.
The Beta regression framework has been extensively studied over the past two decades. For instance, \citet{simas2010} extend Beta regression by modelling both the mean parameter and the precision parameter as linear combinations of covariates, linked via appropriate functions. Similarly, \citet{dasilva2017} incorporate Beta regression into a geographically weighted regression framework, adapting the model to account for spatial variation. Furthermore, \citet{figueroazuniga2013} extend Beta regression by embedding it within a Bayesian hierarchical framework, enabling the inclusion of random effects. The Beta regression model we propose builds on the Bayesian hierarchical framework of \citet{figueroazuniga2013}, while introducing important modifications to the random-effects specification and the variable selection strategy to address the specific challenges of modelling spatio-temporal structures in areal data and selecting from a large set of covariates. Bayesian Beta regression models with spatial dependence have been developed mainly in geostatistical frameworks, where Gaussian spatial random fields enter in the linear predictor to capture distance-based correlation \citep{lagos2017,ParadinasPenninoConesa2018}. For areal data, spatial dependence is more naturally encoded through neighbourhood structures, as in \citet{cepedacuervo2013}, where a spatial autoregressive Beta regression with contiguity-based lags in both mean and dispersion is used. 

To account for spatial and temporal correlation, we incorporate in the linear predictor of the Beta regression model an additive formulation for the random effects. Specifically, we assume an Intrinsic Conditional Auto-Regressive (ICAR) prior for the spatially structured random effect, exploiting its ability to capture structured spatial dependence, which is particularly relevant for regional-level data where geographic proximity strongly influences observed outcomes \citep{banerjee2014, haining2020}. The temporal random effect is modelled as a first-order autoregressive process (AR(1)). Furthermore, in order to capture potential group differences across gender, we include gender-specific intercept terms in the model. In this way, our approach jointly accounts for space, time, and gender influences on regional obesity rates, all of which have been shown to be highly relevant \citep{grimacciarota}.
The additive combination of the spatial and temporal random effect offers a parsimonious and interpretable framework, making it a suitable modelling strategy \citep{banerjee2014}. Although more structured interactions between spatial and temporal components could be incorporated \citep[for a review, see][]{knorrheld2000}, we follow the approach of \citet{waller1997} and \citet{berloco2023}, arguing that prioritizing simplicity and interpretability over added flexible and complexity aligns with the primary objective of this study: to effectively model regional-level patterns in obesity rates while focusing on structured spatio-temporal dynamics.  

With the objective of finding a parsimonious representation of the observed data using a statistical model that is also capable of accurate prediction, we adopt the Stochastic Search Variable Selection (SSVS) approach (\citealp{george1993}, \citealp{george1997}). This method is a Bayesian regularization technique that has proven effective in addressing both prediction and variable selection problems \citep[see][for a detailed discussion]{rockova2012}.
SSVS is particularly attractive because it is well-suited for high-dimensional modelling, provides valid standard errors, and enables the simultaneous estimation of regression coefficients and complexity parameters through computationally efficient Markov chain Monte Carlo (MCMC) techniques.

Although the individual components of our proposed approach, namely Bayesian Beta regression, ICAR spatial effects, AR(1) temporal dynamics, group-specific intercepts, and SSVS, are all well established in the literature, their integration within a single coherent Bayesian hierarchical framework for modeling regional obesity rates is, to our knowledge, novel. Our contribution lies in combining these elements into an areal spatio-temporal model that jointly accounts for spatial and temporal heterogeneity, gender-specific differences, and covariate uncertainty, thereby providing a flexible and comprehensive framework for analysing regional obesity dynamics.

Beyond this methodological integration, we construct a harmonized regional panel dataset on obesity and related covariates for Italy through a systematic process of data collection, validation, and reconciliation across multiple institutional sources. This integrated modelling and data effort enables robust inference and improved prediction of obesity prevalence at the regional level, delivering evidence directly relevant for place-based public health policies.

The paper is structured as follows. In Section \ref{sec:data}, we detail the construction of the dataset, which was assembled from multiple official sources. In Section \ref{sec:methods} we outline the Bayesian hierarchical spatio-temporal Beta regression model (Section \ref{subsec:st_model_likelihood}), providing details about prior assumptions (Section \ref{subsec:st_model_priors}) for the structured spatial and temporal random effects, alongside the gender-specific intercepts. Section \ref{subsec:ssvs} and \ref{subsec: computational_details} discuss the SSVS prior specification for the fixed effects and some computation details, respectively. We then present the results of our analysis in Section \ref{sec:results} and some final considerations in Section \ref{sec:conclusion}. Additional material, including descriptive statistics, posterior summaries, predictive comparisons, geographical cross-validation, and robustness diagnostics, is provided in the appendices.

\section{Data}
\label{sec:data}

The dataset analysed in this study comprises spatio-temporal areal data, which can be conceptualized as a set of time series associated with each spatial unit (i.e., region). 
The dataset is organized as a longitudinal dataset, collecting information on the 20 Italian regions from 2010 to 2022 (13 years).
The Italian regions are divided, according to the NUTS 1 classification\footnote{NUTS classification: https://ec.europa.eu/eurostat/web/nuts} into 5 different geographical zones: North West (NO), North East (NE), Centre (C), South (S) and Islands (SI).
In Table \ref{tab:italian_regions} (Appendix \ref{Appendix1}) we report the regions and the corresponding geographical zone.


In our analysis, a \enquote{weights matrix} $\bm{W}$ is needed to model spatial dependence through the ICAR prior (see Section \ref{sec:methods}). This matrix encodes the neighborhood relationships between regions, and due to its binary nature, a challenge arises in defining neighbors for the two main Italian islands, Sicilia and Sardegna. We treat Sicilia as a neighbor of Calabria, reflecting their geographic proximity across the Strait of Messina. In contrast, Sardegna is modeled as isolated, given its substantial spatial separation from all other regions. This specification was selected after comparing the values of Moran’s index \citep[see][]{banerjee2014} under alternative neighborhood definitions for the two islands 
\citep[see][]{grimacciarota}

The dataset contains aggregated data at the regional level (NUTS 2) with variables grouped into four categories: socio-demography, mortality, lifestyle, and general health-related information (see Table \ref{tab:covariates} in Appendix \ref{Appendix1}). 
We assembled the dataset by extracting information from several official sources: the Health For All platform 
(HFA\footnote{HFA: https://www.istat.it/it/archivio/14562}),
the Benessere Equo e Sostenibile 
(BES\footnote{BES: https://www.istat.it/en/well-being-and-sustainability/the-measurement-of-well-being/bes-report}) report,
 Eurostat databases\footnote{Eurostat: https://ec.europa.eu/eurostat/data/database}
and the archive of the Ministry of Economy and Finance 
(MEF\footnote{MEF archive: www.rgs.mef.gov.it/VERSIONE-I/archivio}).

\begin{figure}[ht]
    \centering
    \includegraphics[width=\textwidth]{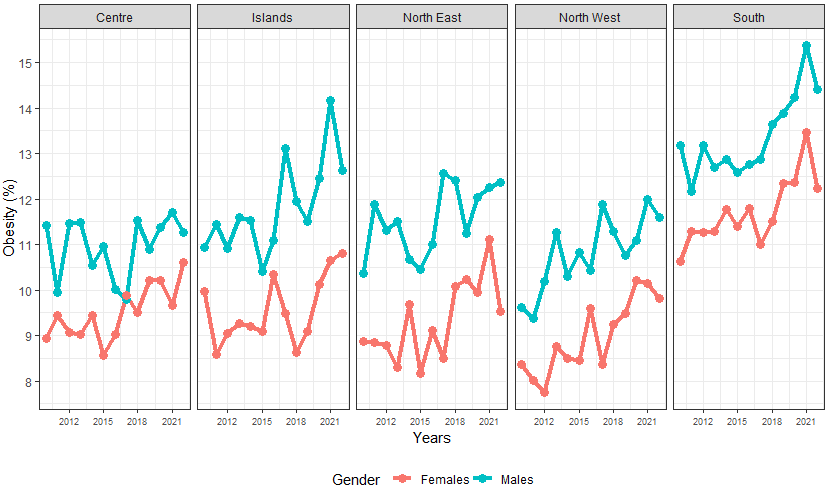}
    \caption{Obesity rate trend from 2010 to 2022 by geographic area and gender.}
\label{fig:ts_obesity}
\end{figure}

The dependent variable is the obesity rate measured yearly for each region and considered separately by gender (source: HFA). We display in Figure \ref{fig:ts_obesity} the obesity trends by geographic area and gender for the time window we consider. 
The definition of obesity follows the WHO classification\footnote{WHO Obesity definition: https://www.who.int/news-room/fact-sheets/detail/obesity-and-overweight}, i.e., having a Body Mass Index (BMI) greater than or equal to 30 defines an obese individual. In the HFA platform, 
obesity rates are official ISTAT sample-based estimates derived from multipurpose household surveys.

For the mortality rate variables in Group 2 of Table \ref{tab:covariates}, which exhibited high multicollinearity and strong linear relationships, we applied a principal component analysis (PCA). These variables represent cause-specific mortality rates, calculated as the number of deaths due to a given cause per 10,000 individuals within the same age and gender group. Among the nine principal components generated, we retained the first two, which together explained more than 70\% of the total variance. The first principal component (PC1) primarily captures diseases related to cardiovascular issues and diabetes, while the second (PC2) reflects factors more associated with stress and mental disorders.
{
Summing up, the final dataset contains 
$p=32$ covariates (see Table \ref{tab:covariates} in Appendix \ref{Appendix1} for more details).}

\section{Methods}\label{sec:methods}

\subsection{Beta regression model}
\label{subsec:st_model_likelihood}

Let \(Y_{its}\) denote the obesity rate in region $i$ $(i=1,\dots,20)$, year \(t\) $(t=1,\dots,13)$ and gender \(s\) $(s=1 \text{ for females and}$ $s=2\text{ for males})$. We specify the following Beta regression model \citep{ferrari2004}:

\begin{equation*}
\begin{aligned}
Y_{its} \mid \mu_{its},\phi \stackrel{ind}{\sim} Beta(\mu_{its} \phi, \ \ (1-\mu_{its}) \phi)
\end{aligned}
\label{likelihood}
\end{equation*}

\begin{equation*}
\begin{aligned}
\ln\left(\frac{\mu_{its}}{1-\mu_{its}}\right) = \eta_{its}
\end{aligned}
\label{likelihood_link}
\end{equation*}

\noindent with the linear predictor \(\eta_{its}\) defined as:
\begin{equation*}
\begin{aligned}
\eta_{its} = \beta_0 + \bm{x}_{its}^T\ \bm{\beta} + \psi_i + \alpha_t + d_s\gamma.
\end{aligned}
\label{lin_predictor}
\end{equation*}
With this formulation we have that $\mu_{its}= E(Y_{its})\in(0,1)$
and \(\text{Var}(Y_{its}) = \left(\mu_{its}(1-\mu_{its})\right)/(1 + \phi) \). It follows that $\mu_{its}$ is the mean obesity rate for region $i$ at year $t$ and for gender $s$, and 
the parameter $\phi$ can be interpreted as a  precision parameter, i.e., given \(\mu_{its}\), the larger the value of \(\phi\), the smaller the variance of \(Y_{its}\). 
The structured space random effect is represented by \(\bm{\psi} = (\psi_1, \psi_2, \ldots , \psi_{20})^T \), while the structured time random effect by
\(\bm{\alpha}\  = (\alpha_1, \alpha_2, \ldots , \alpha_{13})^T \). The global intercept term is $\beta_0$, while the gender-specific intercepts are introduced through the dummy variable $d_s$, which takes the value 0 for females and 1 for males. 

Following \citet{berloco2023}, we adopt an additive specification for the spatial and temporal effects.
More details about the associated prior specification are provided in Section \ref{subsec:st_model_priors}. 
{We denote the covariate vector with \(\bm{x}_{its} = (x_{its1}, x_{its2}, \ldots, x_{itsp})^{T}\) 
and the corresponding fixed effect coefficients with 
\(\bm{\beta} = (\beta_{1}, \beta_{2}, \ldots, \beta_{p})^{T}\).
} 

The Beta parametrization adopted in our model offers distinct advantages. First, by expressing the distribution in terms of its mean \(\mu_{its}\) and precision \(\phi\), the model facilitates a more intuitive interpretation. Second, this formulation simplifies the specification of prior distributions within the Bayesian framework. Prior knowledge can be naturally incorporated through \(\mu_{its}\) via the linear predictor. Simultaneously, the precision parameter \(\phi\) can be assigned a weakly informative prior, allowing the model to flexibly adapt to the data's inherent variability \citep{ferrari2004}.

\subsection{Prior specification}
\label{subsec:st_model_priors}

For the spatially structured random effect \(\bm{\psi}\) we assume an ICAR prior, a standard and computationally convenient specification for structured areal effects, as it encourages local smoothing by borrowing strength across neighboring regions \citep{rue2005,banerjee2014}. Specifically, the spatially structured random effect is modelled as follows: 
\begin{equation*} 
\bm{\psi} \mid \tau_{\psi} \sim \mathcal{N}\big(\bm{0}, [\tau_{\psi}(\bm{D} - \bm{W})]^{-1}\big),
\end{equation*}
where \(\bm{W}\) is a binary symmetric \(20 \times 20\) matrix known as the \enquote{weights matrix} or \enquote{adjacency matrix} \citep[see, e.g.,][]{banerjee2014, haining2020}. The entries of \(\bm{W}\) are defined as follows: \(w_{ij} = 1\) if regions \(i\) and \(j\) share a common boundary (i.e., are first–order contiguous neighbours) and \(w_{ij} = 0\) otherwise. Diagonal elements are set to zero since a region is not considered a neighbour of itself.
The diagonal matrix \(\bm{D}\) is defined as \(\bm{D}=\textbf{Diag}(\bm{W} \cdot \bm{1})\), where 
\(\textbf{Diag}\)
represents the diagonal matrix operator, \(\bm{1}\) denotes a column vector of ones and  
\( \cdot \) denotes the dot product. The diagonal entries $D_{ii}$ represent the number of neighbours for each region $i$.
The scalar \(\tau_{\psi}\) is a precision parameter that scales the degree of spatial smoothing in the model:  higher values of $\tau_{\psi}$ lead to stronger spatial smoothing, while lower values allow for greater heterogeneity across regions.
The full-conditional distribution of the $i$-th spatial random effects \(\psi_i\) is given by:  
\begin{equation}
\psi_i \mid \psi_{j \neq i}, \tau_{\psi} \sim \mathcal{N}\bigg(\sum_{j \sim i} \frac{\psi_j}{D_{ii}}, \frac{1}{D_{ii} \tau_{\psi}}\bigg),
\label{eq:full_conditional_distrib_icar}
\end{equation}  
where \(j \sim i\) indicates that regions \(i\) and \(j\) are neighbours. As shown in Equation \eqref{eq:full_conditional_distrib_icar}, the conditional mean of \(\psi_i\) is the average of the values of its neighbours, while its variance decreases as the number of neighbours increases. 

For what concerns the structured time random effects, following \citet{knorrheld2000} and \citet{berloco2023}, we model the temporal random vector \(\bm{\alpha}\) using a first-order auto-regressive (AR(1)) process with the following structure:
\begin{equation*}
\alpha_1 \mid \tau_{\alpha} \sim \mathcal{N}(0, \tau_{\alpha}^{-1}),
\end{equation*}
\begin{equation*}
\alpha_t \mid \alpha_{t-1}, \rho, \tau_{\alpha} \sim \mathcal{N}(\rho \alpha_{t-1}, \tau_{\alpha}^{-1}), \quad t = 2, \ldots, 13,
\end{equation*}
where \(\tau_{\alpha}\) is the precision parameter and \(\rho\) is the auto-regressive coefficient. We assign a uniform prior to \(\rho\) over the interval $(0,1)$:
\begin{equation*}
\rho \sim \text{Beta}(1, 1),
\label{eq:rho_beta}
\end{equation*}
reflecting a prior belief in positive autocorrelation among the temporal random effects while preventing explosive dynamics in the AR(1) process.

The gender-specific intercepts are modelled in an additive way. First, we recall that the individual \( \psi_i \)'s cannot be uniquely determined. This issue is resolved by imposing a sum to zero constraint $\sum_{i=1}^{n}\psi_i = 0$ \citep{geobugsmanual}  and by modelling the global intercept term \(\beta_0\) as suggested by \cite{besag1995}:  
\begin{equation*}
\begin{aligned}
p(\beta_0) \propto 1, \quad \text{for} \quad \beta_0 \in (-\infty, +\infty),
\end{aligned}
\end{equation*}
indicating a flat, improper prior for \(\beta_0\) over the entire real line.
A gender-specific shift is introduced through the binary indicator variable \(d_s\). The gender effect coefficient $\gamma$ allows the definition of gender-specific intercepts as
\[
\xi_s = \beta_0 + d_s\gamma,
\]
so that \(\xi_1 = \beta_0\) for females and \(\xi_2 = \beta_0 + \gamma\) for males. 
We assign to $\gamma$ a weakly informative normal prior: 
\begin{equation*} 
\gamma \sim \mathcal{N}(0, \sigma^2_\gamma=100).
\end{equation*}

For the structured spatial and temporal random effects, we assign a Gamma prior to the respective precision parameters $\tau_{\psi}$ and $\tau_{\alpha}$. In particular, following the recommendation of \citet{kelsall1999}, for the spatial precision $\tau_{\psi}$ we use a $\text{Gamma}(\alpha_{\tau}, \beta_{\tau})$ prior with $\alpha_{\tau} = 0.5$ and $\beta_{\tau} = 0.0005$; this reflects a conservative and weakly informative belief about spatial dependence \citep{geobugsmanual}. Although originally proposed for spatial effects, we extended this prior to the temporal precision $\tau_{\alpha}$, acknowledging the presence of temporal correlation in our setting, favoring shrinkage while preserving flexibility to capture stronger temporal structure when supported by the data.

Finally, we assign a prior to the Beta precision parameter \( \phi \) following the approach proposed by \citet{figueroazuniga2013}, which extends the prior structure suggested by \citet{gelman2006} specifically for the Beta distribution, by assuming:
\begin{equation*}
\phi = (aB)^2, \quad B \sim \text{Beta}(1+\epsilon, 1+\epsilon),
\end{equation*}
where \( a \) is a large constant (e.g., 50) and \( \epsilon \) is a small constant (e.g., 0.1).

To complete the Bayesian hierarchical formulation, we define the prior for the covariate coefficients $\bm \beta$ using the SSVS approach, as discussed in Section \ref{subsec:ssvs}.

\begin{table}[h]
\caption{Summary of the model parameters.}\label{tab:model_params}%
\begin{tabular}{@{}ll@{}}
\toprule
\textbf{Parameter} & \textbf{Description} \\
\midrule
$\xi_s$ & Gender-specific intercept term ($s=1,2$) \\
$\psi_i$ & Structured spatial random effect ($i=1,\ldots,20$) \\
$\tau_{\psi}$ & Precision parameter for the spatial random effect \\
$\alpha_t$ & Structured temporal random effect ($t=1,\ldots,13$) \\
$\tau_{\alpha}$ & Precision parameter for the temporal random effect \\
$\rho$ & AR(1) coefficient for the temporal random effect \\
$\phi$ & Beta precision parameter \\
$\beta_k$ & Regressor coefficient ($k=1,\ldots,p$) \\
$\sigma_k$, $\omega_k$, $\theta_k$ & Parameters from the SSVS prior \\
\botrule
\end{tabular}
\end{table}

\subsection{Stochastic Search Variable Selection Prior}
\label{subsec:ssvs}
To model the fixed effects coefficients \(\beta_k, k = 1, \dots, p\), we adopt a SSVS prior specification \citep{george1993}. This approach combines a spike-and-slab mixture prior to probabilistically determine covariate inclusion. 
In a nutshell, SSVS enables each regressor to be significantly associated with the response (i.e., coefficient different from zero) or shrunk toward zero. This is achieved by assigning a prior that assumes independence between regressors, with marginals defined as a mixture of two Normal distributions, both centered at zero. 
The spike component corresponds to negligible or irrelevant effects, while the slab component allows for greater variance and captures substantial or practically meaningful associations.
Specifically, we assign the following hierarchical prior structure:
\begin{equation*}
\begin{aligned}
    \beta_k\mid \sigma^2_k &\sim \mathcal{N}(0, \sigma_k^2), \qquad k=1,\ldots,p \\
    \sigma_k^2 \mid \omega_k  & =  (1 - \omega_k) \tau^2 + \omega_k\, c^2 \tau^2, \\
    \omega_k \mid \theta_k &\sim \text{Bernoulli}(\theta_k), \\
    \theta_k &\sim \text{Uniform}(0, 1).
\end{aligned}
\label{eq:ssvs_prior}
\end{equation*}
In this formulation, \(\omega_k\) is a binary latent variable indicating whether the \(k\)-th covariate is \enquote{practically significant} (\(\omega_k = 1\)) or not. The prior probability of inclusion of the \(k\)-th covariate is denoted by \(\theta_k\), to which we assign a Uniform hyper-prior, avoiding imposing strong prior assumptions about the relevance of individual predictors.
The parameters \(\tau^2\) and \(c^2 \tau^2\) represent the variances of the spike and slab components, respectively. 
The intersection points of the spike and slab densities define a threshold \(\zeta\) for practical significance:
\(
\zeta = \pm \tau \sqrt{2c^2\log(c)/(c^2 - 1)},
\)
such that coefficients falling within the interval \([-\zeta, +\zeta]\) are interpreted as practically negligible.
Therefore, in accordance with \citet{rockova2012}, we fix \(\tau\) and \(c\) to constant values  in order to control the separation between negligible and relevant effects (see Section \ref{subsec: computational_details}).
Variable selection is then achieved in practice by either identifying the model with the highest posterior probability, including or excluding covariates based on their marginal posterior inclusion probabilities, or evaluating the extent to which the marginal posterior distributions of the regression coefficients are shrunk toward zero \citep[for further details see][]{rockova2012}.

\subsection{Computational details}
\label{subsec: computational_details}

A summary of all the model parameters is reported in Table \ref{tab:model_params}. To conduct Bayesian inference, we implemented the proposed spatio-temporal hierarchical model using the \texttt{Nimble} package \citep{devalpine2017, devalpine2024}, which enables efficient specification and sampling of complex models. 
To improve convergence of the MCMC algorithm, we normalized the variables using min-max scaling and standardization, as appropriate.

With regard to the SSVS specification, we set the slab scaling factor to \(c = 4000\), and the intersection term to \(\zeta = 0.001\), reflecting the threshold of practical significance identified through iterative exploration and consideration of the magnitude of the coefficients (we refer to Appendix \ref{Appendix4} for further details about the choice and sensitivity analysis of SSVS parameters). This resulted in \(\tau = 0.00025\) (see Section \ref{subsec:ssvs}), a value corresponding to a very small standard deviation for the spike around 0. The slab variance, given by \(\tau^2 c^2\), is approximately equal to 1. This configuration aligns the slab variance with the suggested scale of variation for coefficients under the logit link function, as suggested by \citet{mcelreath2016}.

We ran two parallel chains with overdispersed initial values for a total of 150{,}000 iterations each. 
We discarded an initial burn-in period of 50{,}000 iterations from each chain to ensure convergence to the stationary distribution. To reduce storage requirements and mitigate the effects of residual autocorrelation in the MCMC output, we thinned posterior samples by retaining every 10th iteration. This resulted in a total of 10{,}000 posterior draws used for inference. The total computation time required to fit the model was approximately 30 minutes per MCMC chain on a standard laptop.

We assessed convergence and sampling efficiency using multiple diagnostics, including rank-normalized $\widehat{R}$ statistics and effective sample size (ESS) measures as proposed by \citet{vehtari2021}, together with the more classic ESS and $\widehat{R}$ \citep{coda2006,gelman_rubin_1992, gelman_brooks_1998}. Detailed convergence diagnostics are reported in Appendix \ref{Appendix5}.

\section{Results}
\label{sec:results}

Table \ref{tab:descriptive_gender} presents the posterior summary statistics for the gender-specific intercepts, which are defined as \(\xi_1 = \beta_0\) for females and \(\xi_2 = \beta_0 + \gamma\) for males (see Section \ref{subsec:st_model_priors}). The mean of the intercept for females is \(-1.09\), indicating a lower baseline effect compared to males, whose corresponding mean is \(0.33\). The standard deviation is slightly higher for males (\(0.18\)) than for females (\(0.16\)).
This difference in variability is further reflected in the coefficients of variation (CV), suggesting a slightly higher sampling variability for the male intercept. Both distributions are approximately symmetric. The kurtosis values are close to that of a normal distribution for males (\(3.03\)), whereas the female-specific distribution exhibits slightly heavier tails (\(3.48\)).
\begin{table}[h!]
\centering
\caption{Posterior summary statistics for the gender-specific intercepts $\xi_1$ and $\xi_2$.}
\label{tab:descriptive_gender}
\begin{tabular}{lccccc}
\toprule
\textbf{Parameter} & \textbf{Mean} & \textbf{SD} & \textbf{CV} & \textbf{Skewness} & \textbf{Kurtosis} \\
\midrule
{$\xi_1$} {(Females)} & -1.09   & 0.16   & -0.15   & -0.22   & 3.48 \\ 
{$\xi_2$} {(Males)}   &  0.33   & 0.18   &  0.53   & -0.11   & 3.03 \\ 
\hline
\end{tabular}
\end{table}
\begin{figure}[h!]
    \centering
    \includegraphics[width=0.5\textwidth]{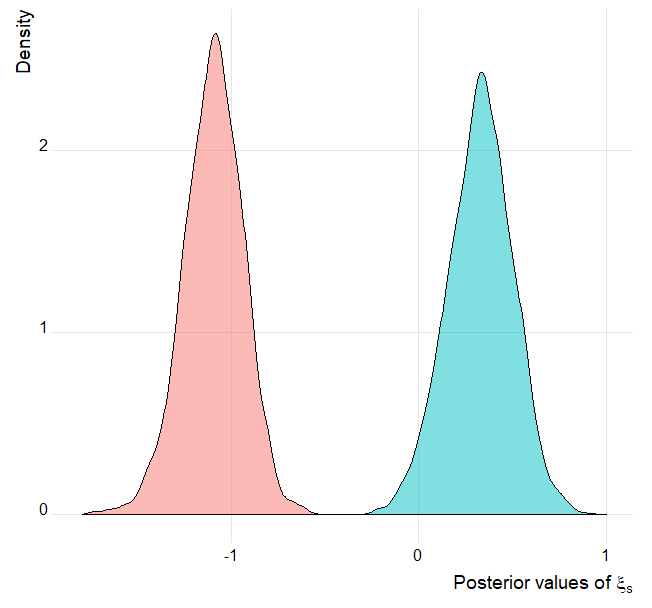}
    \caption{Posterior distributions of the gender-specific intercepts: $\xi_1$ for females (in pink) and $\xi_2$ for males (in light blue)}
    \label{fig:Gender_intercept_post_density}
\end{figure}
Figure \ref{fig:Gender_intercept_post_density} depicts the posterior density estimates of the gender-specific intercepts. The clear separation between the two distributions highlights the substantial contribution of gender-specific effects in explaining differences in obesity prevalence.

\begin{figure}[h!]
    \centering
    \includegraphics[width=0.85\textwidth]{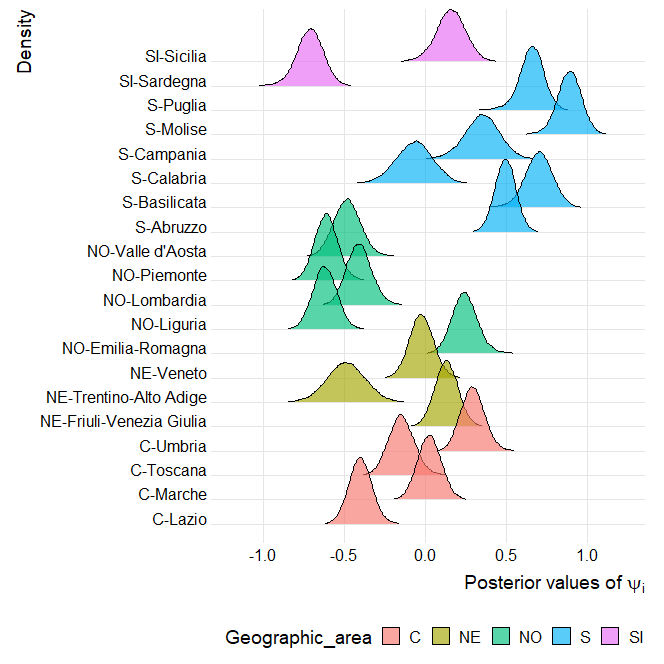}
\caption{Posterior density of the space random effects $\psi_i$, coloured by geographical zone (NO=North West, NE=North East, C=Centre, S=South, SI=Islands, see Table \ref{tab:italian_regions})}
\label{fig:space_RandomEffect_post_density}
\end{figure}

\begin{figure}[h!]
    \centering
    \includegraphics[width=0.7\textwidth]{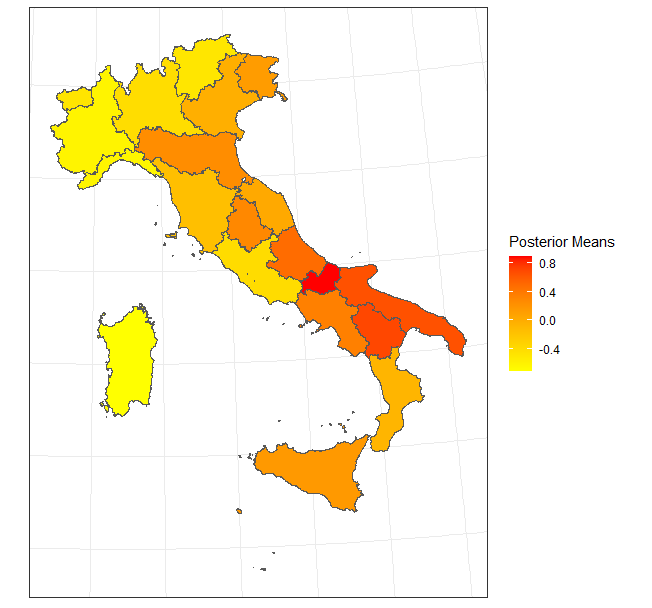}
\caption{Posterior mean values of the spatial random effects $\psi_i$ for each Italian region}
\label{fig:map_RandomEffect_post_density}
\end{figure}

The structured spatial random effects can be visualized through the posterior densities plotted in Figure \ref{fig:space_RandomEffect_post_density}. The plot shows the posterior densities for all regions, highlighting the contribution of regional random effects to the overall obesity estimates according to the model. Overall, the ICAR prior effectively induces a spatial structure that captures regional dependencies while allowing for evident heterogeneity, as reflected in the varying magnitudes of $\psi_i$ across macro areas and individual regions. To complement the density plots, Figure \ref{fig:map_RandomEffect_post_density} provides a map representation of the posterior mean values of the spatial random effects, facilitating the interpretation of their spatial distribution.
While spatial proximity generally corresponds to similar values of the spatial random effect, notable exceptions suggest that additional determinants beyond geography influence regional obesity rates. As shown by the posterior summary statistics in Table \ref{tab:tab:psi_summary_with_regions} of Appendix \ref{Appendix2}, regions with negative posterior mean values, such as Piemonte (-0.61), Liguria (-0.62), Trentino-Alto Adige (-0.49), and Sardegna (-0.71, the lowest), are associated with lower obesity rates. In contrast, with the exception of Calabria, all southern regions, particularly Molise (0.89, the highest), as well as Umbria (0.29), exhibit strongly positive spatial effects. 
The uncertainty in spatial effects varies across regions, as reflected in the standard deviations (SD) of the posterior distributions. Trentino-Alto Adige and Calabria (SD = 0.12, the highest), Campania, and Sicilia exhibit greater variability, whereas Veneto, Friuli-Venezia Giulia, Umbria, Marche, Lazio and Abruzzo show more spiked posterior estimates, with Abruzzo having the lowest SD (0.06). Additionally, the distributions reveal a skewness pattern: northern regions such as Piemonte, Lombardia, Liguria, and Lazio exhibit right-skewness, while southern regions generally show left-skewness. Notably, all the regions have kurtosis values exceeding 3, suggesting pronounced posterior tails, with Puglia being the region with the heaviest tails.
To summarize, while spatial correlation in obesity rates is evident, the variation across regions is not as smooth as one might expect from a prior ICAR specification. Some regional posterior densities deviate strongly from those of neighbouring regions, suggesting localized spatial heterogeneity that is not entirely captured by the prior smoothing assumptions. As regards the posterior summary statistics for the spatial precision parameter $\tau_{\psi}$, we refer the reader to Appendix \ref{Appendix2}.

\begin{figure}[h!]
    \centering
    \includegraphics[width=0.8\textwidth]{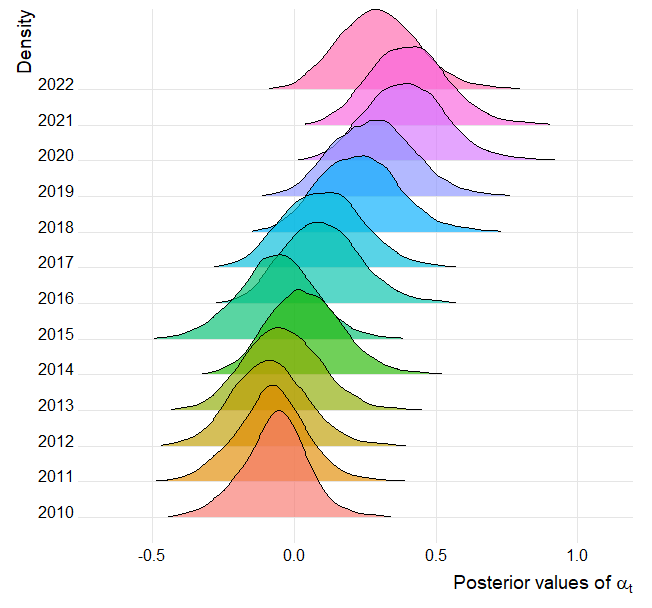}
\caption{Posterior density of the time random effects $\alpha_t$, coloured by year}
\label{fig:time_RandEff_post_density}
\end{figure}

Posterior density estimates of the time-specific random effects are shown in Figure \ref{fig:time_RandEff_post_density}. A clear temporal pattern emerges: during the first half of the observational window (2010–2015), the random effects remain relatively stable and are centered around values slightly below zero. In contrast, the second half of the period exhibits a marked upward shift in the magnitude of the effects.
Between 2010 and 2015, the posterior probability that the temporal effects exceed zero remains relatively stable, fluctuating around 0.3, with the exception of 2014 which shows a modest increase. Starting in 2016, these probabilities rise significantly: both 2016 and 2017 reach approximately 0.76, and from 2018 onward, the vast majority (if not all) of posterior samples are positive. These trends are summarized in Table \ref{tab:alpha_positive_proportions}.

\begin{table}[!h]
\centering
\caption{\label{tab:alpha_positive_proportions} 
Posterior probability that the temporal effect $\alpha_t$ exceed zero (PP0).}
\renewcommand{\arraystretch}{1.2}
\setlength{\tabcolsep}{4pt}
\begin{tabular}{lccccccccccccc}
\toprule
$t$ & 1 & 2 & 3 & 4 & 5 & 6 & 7 & 8 & 9 & 10 & 11 & 12 & 13 \\
Year & 2010 & 2011 & 2012 & 2013 & 2014 & 2015 & 2016 & 2017 & 2018 & 2019 & 2020 & 2021 & 2022 \\
\midrule
PP0 & 
0.28 & 0.25 & 0.25 & 0.38 & 0.61 & 0.32 & 0.76 & 0.76 & 0.96 & 0.98 & 1.00 & 1.00 & 0.99 \\
\bottomrule
\end{tabular}
\end{table}

As confirmed by the posterior summary statistics shown in Table \ref{tab:tab:alpha_summary_with_time} of Appendix \ref{Appendix2}, the mean values of the $\alpha_t$ parameters range from negative values, such as $\alpha_2$ and $\alpha_3$, which have the lowest mean (-0.08), to positive values, with $\alpha_{12}$ exhibiting the highest one (0.41). These differences highlight the varying influence of the time random effects across different time points. While some periods exhibit negative random effects, indicating a lower-than-expected influence on the obesity rates, others demonstrate positive effects, which may indicate periods of higher-than-expected influence.
The standard deviations of the $\alpha_t$ parameters are relatively stable, ranging from 0.12 to 0.15. This consistency suggests that while the magnitude of time random effects may differ across time points, the variability within each time point is relatively homogeneous. 

\begin{table}[h!]
\centering
\caption{Posterior summary statistics for the auto-regressive coefficient $\rho$.}
\label{tab:descriptive_rho}
\begin{tabular}{lccccc}
\toprule
\textbf{Parameter} & \textbf{Mean} & \textbf{SD} & \textbf{CV} & \textbf{Skewness} & \textbf{Kurtosis} \\
\midrule
{$\rho$} & 0.85 & 0.14 & 0.16 & -1.48 & 5.62\\
\bottomrule
\end{tabular}
\end{table}

The autoregressive parameter $\rho$ exhibits a negatively skewed posterior distribution, with a posterior mean of 0.85 (Table \ref{tab:descriptive_rho}), indicating strong temporal persistence in regional obesity rates. This suggests that obesity levels in a given year are highly correlated with those in previous years, resulting in a smooth temporal evolution rather than abrupt year-to-year changes. Such persistence is coherent with the influence of slowly evolving underlying factors, rather than short-term shocks. Although broader socioeconomic, behavioral, or epidemiological processes may have contributed to this upward trend, the aggregated nature of the data and the lack of information on specific policies or interventions prevent a more detailed analysis. We therefore interpret the estimated value of $\rho$ primarily as capturing long-run temporal dependence in obesity dynamics, rather than effects associated with specific policy actions or events.
As regards the posterior summary statistics for the temporal precision parameter $\tau_{\alpha}$, we refer the reader to Table \ref{tab:descriptive_hyperpriors} in Appendix \ref{Appendix2}.

\begin{table}[h!]
\centering
\caption{Posterior summary statistics for the Beta precision parameter $\phi$.}
\label{tab:descriptive_phi}
\begin{tabular}{lccccc}
\toprule
\textbf{Parameter} & \textbf{Mean} & \textbf{SD} & \textbf{CV} & \textbf{Skewness} & \textbf{Kurtosis} \\
\midrule
$\phi$ & 37.67 & 2.39 & 0.06 & 0.14 & 3.00 \\
\botrule
\end{tabular}
\end{table}

The posterior summary statistics of the Beta precision parameter $\phi$ reported in Table \ref{tab:descriptive_phi} are not straightforward to interpret, primarily because they depend on the scale of the data. The parameter becomes more interesting when compared between models: a model with a higher posterior precision parameter suggests a better ability to explain the variability of the Beta-distributed response variable, due to the relation that links the variance to the precision $\phi$ (see Section \ref{sec:methods}). 
We just highlight that the Bayesian updating process was very pronounced, as the prior distribution, which spans a wide range (between $0$ and $2500$), is effectively updated to a more concentrated range around the mean of 37.67. The reduction in uncertainty is very high, as reflected by the low coefficient of variation equal to 0.06. 

\begin{figure}[h]
    \centering
    \includegraphics[width=0.85\textwidth]{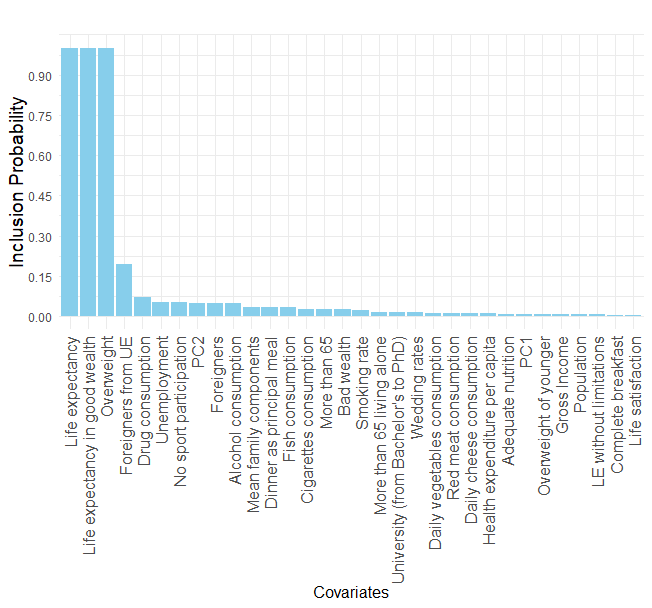}
\caption{Posterior probability evaluation of the inclusion parameters $\omega_k$ for all the covariates (for the description of the covariates see Table \ref{tab:covariates} in Appendix \ref{Appendix1})}
\label{fig:ssvs_gamma_post_evaluation}
\end{figure}

As discussed in Section \ref{subsec:ssvs}, we adopt the SSVS approach to perform inference on the fixed-effect coefficients $\beta_k$. Consequently, the evaluation of their posterior distribution is intrinsically linked to the variable selection induced by the spike and slab prior. The SSVS hierarchical prior specification can make inference on the $\beta_k$ coefficients challenging, particularly when their posterior distributions are bi-modal, with samples alternating between values near zero and substantially different from zero. This complicates the inferential process. However, we point out that the primary objective of this work is to investigate the spatio-temporal dynamics of obesity rates across Italian regions, rather than exploring or assessing causal relationships between the covariates and the response variable. The inclusion of these covariates is primarily to control for any additional effect that may arise along with the random effects, which are of greater interest in this study. Thus we do not conduct detailed posterior evaluation of the $\beta_k$ coefficients. 
Nevertheless, given the valuable insights offered by the SSVS approach, we perform a brief analysis of the inclusion probabilities to assess which covariates are associated with obesity, after accounting for spatial and temporal random effects, as well as gender. Specifically, the $\omega_k$ parameter plays a central role in determining whether the $k$-th covariate should be included in the model, based on the probability parameter $\theta_k$ (see Section \ref{subsec:ssvs}). By counting the proportion of MCMC iterations in which $\omega_k=1$, we obtain an empirical inclusion probability for each covariate. Figure \ref{fig:ssvs_gamma_post_evaluation} presents these posterior inclusion probabilities. The only covariates that are consistently included in the model are ``life expectancy'', ``life expectancy in good health'' and  ``overweight'' (see Table~\ref{tab:covariates}). All the remaining covariates exhibit very low posterior inclusion probabilities.
This suggests that once the random effects and gender-specific intercepts are incorporated, only three covariates need to be included according to our model specification. The other covariates do not prove to be related to obesity rates and only add unnecessary complexity to the model.
We observe that this is consistent with the structure of the model. In particular, the spatial and temporal random effects act as latent proxies for a wide range of unobserved regional characteristics, such as cultural habits, health system features, environmental conditions, socioeconomic context, or persistent lifestyle patterns. As a consequence, much of the cross-regional variability that might otherwise be captured by observable covariates is absorbed by the spatial and temporal terms, reducing the marginal contribution of many individual predictors. This highlights that, at the regional scale, unobserved or hard-to-measure contextual factors play a dominant role relative to the socioeconomic covariates available in the dataset.

To assess the sensitivity of the SSVS procedure to hyperparameter choices, we conducted a sensitivity analysis to determine whether different SSVS settings yield distinct shrinkage patterns. The full analysis is reported in Appendix \ref{Appendix4}. Briefly, we tested four alternative configurations based on different intersections between the spike and slab densities (see Section \ref{subsec:ssvs}). While posterior inclusion probabilities vary slightly across specifications, the same three covariates consistently emerge as the only ones not shrunk toward zero. All the remaining covariates have posterior inclusion probabilities below the commonly used 0.5 threshold, with a minor exception in one specification (see Appendix \ref{Appendix4} for further details).

\subsection{Predictive performance}
In order to assess the performance of the modelling approach we propose, we considered three alternative model specifications alongside the main model presented in Section \ref{sec:methods}. These variants differ in two aspects: (i) the treatment of the gender effect, modelled either as a fixed or a random effect; and (ii) the prior distribution on the auto-regressive coefficient \(\rho\), specified either as a Beta\((1,1)\) distribution (constraining \(\rho\) to the unit interval) or as a standard Normal distribution (allowing unrestricted support). This results in four model configurations, named M1, M2, M3 and M4, as detailed in Table \ref{tab:model_variants} in Appendix \ref{subsec:2022_xvalid}. Note that M1 (i.e., gender as fixed effect and Beta prior for $\rho$) corresponds to our main model described in Section \ref{sec:methods}. To evaluate and compare the predictive performance of the four models, we implemented both temporal and spatial validation strategies. 
First, we conducted a temporal train–test split, using data from 2022 (\( t = 13 \)) as the test set and data from the preceding years (\( t = 1,\dots,12 \)) for model training (see Appendix \ref{subsec:2022_xvalid}).
Second, we performed a spatial cross-validation procedure by leaving out one representative region from each macro-area: Lombardia (North), Toscana (Centre), and Campania (South). Further details are provided in Appendix \ref{subsec:spatial_xvalid}.
Models were assessed using standard predictive metrics. For what concerns the temporal test, the results reported in Table \ref{tab:evaluation_metrics} indicate that M1 outperforms the competing models in terms of predictive accuracy. The latter is likely due to the temporal dynamics allowed by the Normal prior, which can lead to explosive behaviours and overfitting. The random gender effect offered no improvement over the fixed-effect specifications. Overall, M1 offered the best balance between parsimony and predictive accuracy. 
For what concerns the spatial cross-validation, we obtained smaller performance differences across model specifications than those observed in the temporal validation exercise. This indicates that, once trained on a sufficiently broad set of regions, all four models are able to capture the main spatial patterns in obesity prevalence with comparable accuracy.

\section{Concluding remarks}\label{sec:conclusion}

This study proposes a novel Bayesian hierarchical Beta regression model for analysing the spatio-temporal dynamics of obesity rates across Italian regions from 2010 to 2022. The results highlight pronounced gender differences, regional heterogeneity and dependence, and clear temporal trends, suggesting the importance of spatial, temporal, and demographic factors in explaining obesity patterns.

The choice of the Beta distribution is motivated not only by its support on the unit interval - making it particularly suitable for modelling rates and proportions - but also by its flexibility, as it accommodates a wide range of distributional shapes, depending on the parameter values \citep{johnson1995}. We mention that beyond Beta regression, alternative methodologies for analysing rate or proportion data have been proposed. For example, compositional data analysis has been used to jointly model multiple outcome categories, such as underweight, overweight, and obesity, through log-ratio transformations \citep{mills2008}. While such approaches are valuable when the primary interest lies in the joint dynamics of multiple components, our focus is on modelling a single rate outcome over space and time.
In the present case study, the proposed Bayesian hierarchical Beta regression model incorporates structured spatial and temporal random effects additively, along with gender-specific intercepts. This model proves effective in capturing regional variations and temporal dependencies in obesity rates. While the individual components of the proposed model are well established, their integration into a unified Bayesian hierarchical framework constitutes the main methodological contribution of this work, offering a coherent and flexible approach for analysing regional obesity rates and their spatio-temporal dynamics.

Some methodological limitations warrant consideration. The ICAR prior for the structured spatial random effects assumes complete spatial dependence, which may not fully reflect the underlying spatial processes. Additionally, the additive specification of spatial and temporal random effects, with separate priors, may overlook potential space-time interactions. The exclusion of unstructured spatio-temporal variability, which could increase flexibility in capturing localized deviations, represents a further limitation. Furthermore, although the regional-level analysis improves measurement reliability and mitigates data gaps, it may obscure finer-scale variations at the provincial level. 
Finally, the adjacency matrix assigns equal weight only to immediately neighbouring regions and therefore does not account for distance decay or other forms of connectivity, such as economic or cultural links. While this is a standard and interpretable choice for areal data, future work could explore more flexible weighting schemes to better capture complex spatial relationships.
Despite these limitations, the proposed framework remains parsimonious, interpretable, and effective in predictive terms. Its main strength lies in balancing modelling flexibility with methodological clarity, thereby providing a practically useful approach for the spatio-temporal analysis of regional obesity rates in public health applications.

Our findings highlight the substantial role of the gender-specific random effects in estimating obesity outcomes, underscoring gender as a critical determinant of regional obesity rates. Spatial random effects reveal homogeneous patterns across the North-West, with the exception of Emilia-Romagna, which exhibits higher values. In the North-East, Friuli-Venezia Giulia and Veneto show similar values, while Trentino-Alto Adige displays lower magnitudes with a greater range. The Centre geographical zone is characterized by marked heterogeneity, with Lazio showing significantly lower values and Umbria showing higher ones. Southern regions generally exhibit higher spatial random effects, except for Calabria. The Islands show distinct spatial structures, an aspect enforced by the fact that Sicilia is treated as a neighbour to Calabria in the ICAR prior, while Sardegna is considered spatially isolated.
Temporal random effects indicate a stable trend in the early years, followed by an increase in later years, suggesting a growing upward trend in obesity rates. Under the SSVS prior specification, most coefficients are shrunk toward zero, with only a few covariates being identified as having a relevant association with obesity, namely \enquote{life expectancy}, \enquote{life expectancy in good health}, and \enquote{overweight}. 
This result suggests that spatial and temporal random effects explain a substantial portion of the variability in obesity rates.
Clearly, additional factors, such as environmental characteristics, food environment indicators, or policy-related variables, may play a relevant role but could not be considered due to the limited availability of regional-level data over the study period. The integration of richer covariate information, therefore, represents a natural and promising direction for future research as more comprehensive data become accessible. 

While this study does not aim to establish causal relationships, it highlights the importance of modelling obesity trends through demographic and spatio-temporal structures rather than relying solely on fixed socioeconomic or lifestyle predictors. The results further support the value of incorporating spatio-temporal dependence in public health research.

Future research could extend this framework by considering finer spatial resolutions, such as provinces or municipalities, incorporating individual-level or environmental information, and allowing for spatio-temporal interactions. Alternative prior specifications could also be explored to increase flexibility in the presence of outliers or asymmetric variation. More broadly, the proposed Bayesian hierarchical Beta regression model may be useful for other areal outcomes bounded between 0 and 1 and characterized by spatial and temporal dependence, including applications in public health and the social sciences.

\backmatter

\bibliography{sn-bibliography}

\section*{Data and code availability}
The assembled dataset and the code used for implementing the analysis and running the models are available at the following GitHub repository:\\ 
\url{https://anonymous.4open.science/r/Spatio-Temporal-Dynamics-of-Obesity-in-Italian-Regions-7893/}


\section*{Acknowledgements}
Blinded for review

\section*{Author contribution}
Blinded for review

\section*{Funding}
Not applicable

\section*{Competing interests}
Not applicable

\section*{Ethics approval and consent to participate}
Not applicable 

\section*{Consent for publication}
Not  applicable 

\mycomment{
\section*{Declarations}

\begin{itemize}
\item Funding
\item Conflict of interest/Competing interests (check journal-specific guidelines for which heading to use)
\item Ethics approval and consent to participate
\item Consent for publication
\item Data availability 
\item Materials availability
\item Code availability 
\item Author contribution
\end{itemize}

If any of the sections are not relevant to your manuscript, please include the heading and write `Not applicable' for that section. 

\bigskip
\begin{flushleft}%
Editorial Policies for:

\bigskip\noindent
Springer journals and proceedings: \url{https://www.springer.com/gp/editorial-policies}

\bigskip\noindent
Nature Portfolio journals: \url{https://www.nature.com/nature-research/editorial-policies}

\bigskip\noindent
\textit{Scientific Reports}: \url{https://www.nature.com/srep/journal-policies/editorial-policies}

\bigskip\noindent
BMC journals: \url{https://www.biomedcentral.com/getpublished/editorial-policies}
\end{flushleft}      
}
\newpage
\appendix

\section{Appendix Organization} \label{Appendix1}
We provide additional material in the appendices: Appendix \ref{Appendix1} contains a table reporting the classification of Italy into geographical zones and a table describing the covariates used in the analysis; Appendix \ref{Appendix2} reports additional posterior summary statistics not included in the main text for brevity, Appendix \ref{Appendix3} presents the results of the predictive comparison of some variants of the main model, Appendix \ref{Appendix4} includes sensitivity analyses for different SSVS hyperparameter specifications and, finally, Appendix \ref{Appendix5} provides extensive diagnostics and performance metrics.

\section{Geographical Areas and Covariates} \label{Appendix1}

\begin{table}[h!]
\begin{center}
\begin{minipage}{0.75\textwidth}
\caption{Italian regions grouped by geographical zone.
Source: Geographical classification of Italian regions according to the Italian National Institute of Statistics (ISTAT).
\label{tab:italian_regions}}%
\begin{tabular}{@{}ll@{}}
\toprule
\textbf{Zone} & \textbf{Regions} \\
\midrule
North West (NO) & Piemonte, Valle d'Aosta, Lombardia, Liguria, Emilia-Romagna \\
North East (NE) & Trentino-Alto Adige, Veneto, Friuli-Venezia Giulia \\
Centre (C) & Toscana, Umbria, Marche, Lazio \\
South (S) & Abruzzo, Molise, Campania, Puglia, Basilicata, Calabria \\
Islands (SI) & Sicilia, Sardegna \\
\botrule
\end{tabular}
\end{minipage}
\end{center}
\end{table}

\FloatBarrier

\begin{table}[h!]
\caption{Covariates used in the analysis, grouped into four thematic categories: Socio-Demography, Mortality, Lifestyle, and General Wealth-related Information. The table reports the unit of measurement, availability by gender, and data source for each variable.\label{tab:covariates}}%
\begin{tabular}{@{}p{5cm}p{3.5cm}p{2cm}p{2cm}@{}}
\toprule
\textbf{Variable} & \textbf{Unit of measure} & \textbf{By gender} & \textbf{Source} \\
\midrule
\multicolumn{4}{@{}c@{}}{\textbf{Group 1 – Socio-Demography}} \\
Population & count & yes & Eurostat \\
More than 65 years old & percentage & yes & HFA \\
Foreigners & percentage & yes & HFA \\
Foreigners from EU & percentage & yes & HFA \\
More than 65 y/o living alone & percentage & yes & HFA \\
Mean family components & average & no & HFA \\
Unemployment & percentage & yes & Eurostat \\
Gross Income & euro (thousands) & no & Eurostat \\
Wedding rates & percentage & no & HFA \\
University (from Bachelor's to PhD) & percentage & yes & Eurostat \\
\midrule
\multicolumn{4}{@{}c@{}}{\textbf{Group 2 – Mortality}} \\
Cancer of digestive system & percentage & yes & HFA \\
Cancer of stomach & percentage & yes & HFA \\
Diabetes & percentage & yes & HFA \\
Mental disorders & percentage & yes & HFA \\
Blood diseases & percentage & yes & HFA \\
Heart diseases & percentage & yes & HFA \\
Digestive diseases & percentage & yes & HFA \\
Liver diseases & percentage & yes & HFA \\
Suicide & percentage & yes & HFA \\
\midrule
\multicolumn{4}{@{}c@{}}{\textbf{Group 3 – Lifestyle}} \\
Overweight & percentage & yes & HFA \\
Overweight of younger & percentage & yes & HFA \\
Cigarette consumption & count & yes & HFA \\
Smoking rate & percentage & yes & BES \\
Complete breakfast & percentage & yes & HFA \\
Daily cheese consumption & percentage & yes & HFA \\
Red meat consumption & percentage & yes & HFA \\
Fish consumption & percentage & yes & HFA \\
Daily vegetables consumption & percentage & yes & HFA \\
Dinner as principal meal & percentage & yes & HFA \\
No sport participation & percentage & yes & HFA \\
Adequate nutrition & percentage & yes & BES \\
Alcohol consumption & percentage & yes & BES \\
\midrule
\multicolumn{4}{@{}c@{}}{\textbf{Group 4 – General Wealth-related Information}} \\
Bad wealth & percentage & yes & HFA \\
Life expectancy (LE) & years & yes & HFA \\
LE in good wealth & years & yes & HFA \\
LE without limitations & years & yes & HFA \\
Drug consumption & percentage & yes & HFA \\
Life satisfaction & percentage & yes & BES \\
Health expenditure per capita & euro/population & yes & MEF \\
\botrule
\end{tabular}
\footnotetext{Source: HFA = Health for All database; BES = Italian Equitable and Sustainable Well-being report; MEF = Italian Ministry of Economy and Finance; Eurostat = Statistical Office of the European Union.}
\end{table}

\newpage
\section{Posterior Summary Statistics}\label{Appendix2}

This appendix provides the parameter posterior summary statistics not reported in the main text.

\FloatBarrier
\begin{table}[!h]
\centering
\caption{\label{tab:tab:psi_summary_with_regions}Posterior summary statistics for region-specific parameters $\psi_1$ to $\psi_{20}$.}
\centering
\fontsize{9}{11}\selectfont
\begin{tabular}[t]{lccccc}
\toprule
\textbf{Parameter (Region)} & \textbf{Mean} & \textbf{SD} & \textbf{CV} & \textbf{Skewness} & \textbf{Kurtosis} \\
\midrule
\cellcolor{gray!10}{{$\psi_{1}$} (Piemonte)} & \cellcolor{gray!10}{-0.61} & \cellcolor{gray!10}{0.07} & \cellcolor{gray!10}{-0.12} & \cellcolor{gray!10}{0.18} & \cellcolor{gray!10}{3.46}\\
{$\psi_{2}$} (Valle d'Aosta) & -0.48 & 0.09 & -0.18 & 0.34 & 3.96\\
\cellcolor{gray!10}{{$\psi_{3}$} (Lombardia)} & \cellcolor{gray!10}{-0.41} & \cellcolor{gray!10}{0.08} & \cellcolor{gray!10}{-0.19} & \cellcolor{gray!10}{0.25} & \cellcolor{gray!10}{3.41}\\
{$\psi_{4}$} (Liguria) & -0.62 & 0.08 & -0.12 & 0.11 & 3.42\\
\cellcolor{gray!10}{{$\psi_{5}$} (Emilia-Romagna)} & \cellcolor{gray!10}{0.25} & \cellcolor{gray!10}{0.08} & \cellcolor{gray!10}{0.33} & \cellcolor{gray!10}{0.32} & \cellcolor{gray!10}{3.59}\\
{$\psi_{6}$} (Trentino-Alto Adige) & -0.49 & 0.12 & -0.25 & 0.02 & 3.19\\
\cellcolor{gray!10}{{$\psi_{7}$} (Veneto)} & \cellcolor{gray!10}{-0.02} & \cellcolor{gray!10}{0.07} & \cellcolor{gray!10}{-3.34} & \cellcolor{gray!10}{0.06} & \cellcolor{gray!10}{3.13}\\
{$\psi_{8}$} (Friuli-Venezia Giulia) & 0.13 & 0.07 & 0.55 & 0.03 & 3.08\\
\cellcolor{gray!10}{{$\psi_{9}$} (Toscana)} & \cellcolor{gray!10}{-0.15} & \cellcolor{gray!10}{0.08} & \cellcolor{gray!10}{-0.56} & \cellcolor{gray!10}{0.18} & \cellcolor{gray!10}{3.30}\\
{$\psi_{10}$} (Umbria) & 0.29 & 0.07 & 0.26 & 0.19 & 3.29\\
\cellcolor{gray!10}{{$\psi_{11}$} (Marche)} & \cellcolor{gray!10}{0.03} & \cellcolor{gray!10}{0.07} & \cellcolor{gray!10}{2.73} & \cellcolor{gray!10}{0.03} & \cellcolor{gray!10}{3.09}\\
{$\psi_{12}$} (Lazio) & -0.40 & 0.07 & -0.18 & 0.20 & 3.55\\
\cellcolor{gray!10}{{$\psi_{13}$} (Abruzzo)} & \cellcolor{gray!10}{0.49} & \cellcolor{gray!10}{0.06} & \cellcolor{gray!10}{0.13} & \cellcolor{gray!10}{-0.03} & \cellcolor{gray!10}{3.05}\\
{$\psi_{14}$} (Molise) & 0.89 & 0.08 & 0.09 & -0.25 & 3.57\\
\cellcolor{gray!10}{{$\psi_{15}$} (Campania)} & \cellcolor{gray!10}{0.35} & \cellcolor{gray!10}{0.11} & \cellcolor{gray!10}{0.32} & \cellcolor{gray!10}{-0.31} & \cellcolor{gray!10}{3.65}\\
{$\psi_{16}$} (Puglia) & 0.65 & 0.08 & 0.12 & -0.55 & 4.34\\
\cellcolor{gray!10}{{$\psi_{17}$} (Basilicata)} & \cellcolor{gray!10}{0.70} & \cellcolor{gray!10}{0.09} & \cellcolor{gray!10}{0.13} & \cellcolor{gray!10}{-0.26} & \cellcolor{gray!10}{3.58}\\
{$\psi_{18}$} (Calabria) & -0.07 & 0.12 & -1.69 & -0.13 & 3.12\\
\cellcolor{gray!10}{{$\psi_{19}$} (Sicilia)} & \cellcolor{gray!10}{0.16} & \cellcolor{gray!10}{0.10} & \cellcolor{gray!10}{0.60} & \cellcolor{gray!10}{-0.22} & \cellcolor{gray!10}{3.60}\\
{$\psi_{20}$} (Sardegna) & -0.71 & 0.09 & -0.12 & -0.32 & 3.79\\
\bottomrule
\end{tabular}
\end{table}

\begin{table}[!h]
\centering
\caption{\label{tab:tab:alpha_summary_with_time}Posterior summary statistics for time-specific parameters $\alpha_1$ to $\alpha_{13}$.}
\centering
\fontsize{9}{11}\selectfont
\begin{tabular}[t]{lccccc}
\toprule
\textbf{Parameter (Year} & \textbf{Mean} & \textbf{SD} & \textbf{CV} & \textbf{Skewness} & \textbf{Kurtosis} \\
\midrule
\cellcolor{gray!10}{{$\alpha_{1}$} (2010)} & \cellcolor{gray!10}{-0.07} & \cellcolor{gray!10}{0.12} & \cellcolor{gray!10}{-1.83} & \cellcolor{gray!10}{-0.02} & \cellcolor{gray!10}{3.79}\\
{$\alpha_{2}$} (2011) & -0.08 & 0.13 & -1.65 & 0.11 & 3.64\\
\cellcolor{gray!10}{{$\alpha_{3}$} (2012)} & \cellcolor{gray!10}{-0.08} & \cellcolor{gray!10}{0.14} & \cellcolor{gray!10}{-1.63} & \cellcolor{gray!10}{0.29} & \cellcolor{gray!10}{3.54}\\
{$\alpha_{4}$} (2013) & -0.04 & 0.14 & -3.51 & 0.26 & 3.36\\
\cellcolor{gray!10}{{$\alpha_{5}$} (2014)} & \cellcolor{gray!10}{0.04} & \cellcolor{gray!10}{0.14} & \cellcolor{gray!10}{3.13} & \cellcolor{gray!10}{0.37} & \cellcolor{gray!10}{3.44}\\
{$\alpha_{6}$} (2015) & -0.06 & 0.14 & -2.24 & 0.01 & 3.31\\
\cellcolor{gray!10}{{$\alpha_{7}$} (2016)} & \cellcolor{gray!10}{0.10} & \cellcolor{gray!10}{0.14} & \cellcolor{gray!10}{1.38} & \cellcolor{gray!10}{0.36} & \cellcolor{gray!10}{3.31}\\
{$\alpha_{8}$} (2017) & 0.11 & 0.14 & 1.36 & 0.23 & 3.09\\
\cellcolor{gray!10}{{$\alpha_{9}$} (2018)} & \cellcolor{gray!10}{0.23} & \cellcolor{gray!10}{0.14} & \cellcolor{gray!10}{0.62} & \cellcolor{gray!10}{0.32} & \cellcolor{gray!10}{3.25}\\
{$\alpha_{10}$} (2019) & 0.28 & 0.15 & 0.52 & 0.26 & 3.14\\
\cellcolor{gray!10}{{$\alpha_{11}$} (2020)} & \cellcolor{gray!10}{0.40} & \cellcolor{gray!10}{0.15} & \cellcolor{gray!10}{0.37} & \cellcolor{gray!10}{0.36} & \cellcolor{gray!10}{3.41}\\
{$\alpha_{12}$} (2021) & 0.41 & 0.14 & 0.35 & 0.40 & 3.48\\
\cellcolor{gray!10}{{$\alpha_{13}$} (2022)} & \cellcolor{gray!10}{0.30} & \cellcolor{gray!10}{0.14} & \cellcolor{gray!10}{0.48} & \cellcolor{gray!10}{0.31} & \cellcolor{gray!10}{3.27}\\
\bottomrule
\end{tabular}
\end{table}

\begin{table}[h!]
\centering
\caption{Posterior summary statistics for the hyperparameters of space ($\tau_{\psi}$), and time ($\tau_{\alpha}$) random effects}
\label{tab:descriptive_hyperpriors}
\begin{tabular}{lccccc}
\hline
\textbf{Parameter} & \textbf{Mean} & \textbf{SD} & \textbf{CV} & \textbf{Skewness} & \textbf{Kurtosis} \\
\hline
$\tau_{\psi}$ (Space precision) & 2.37 & 0.82 & 0.34 & 0.80 & 4.14 \\
$\tau_{\alpha}$ (Time precision) & 100.84 & 65.26 & 0.65 & 2.19 & 12.08 \\
\hline
\end{tabular}
\end{table}

Focusing on the precision parameters for the spatial and temporal random effects ($\tau_{\psi}$ and $\tau_{\alpha}$, respectively), we observe substantial Bayesian updating, particularly for the temporal component (see Table \ref{tab:descriptive_hyperpriors}). Beginning with conservative Gamma priors, the posterior distribution of the spatial precision $\tau_{\psi}$ exhibits only a moderate increase. This restrained growth can be partly attributed to the ICAR prior structure, in which the full-conditional precision is scaled by the number of neighboring regions.
In contrast, the temporal precision $\tau_{\alpha}$, whose prior specification does not involve such a multiplicative adjustment, undergoes a more pronounced shift, with a posterior mean markedly higher than that of $\tau_{\psi}$. However, direct comparison between these two parameters should be approached with caution, as their prior formulations differ substantially, rendering such comparisons potentially misleading.
Both posterior distributions are characterized by high skewness and kurtosis, suggesting strong positive asymmetry and heavy tails. 

\section{Predictive Performance Comparison}\label{Appendix3}
\subsection{Predictive Test for Year 2022}\label{subsec:2022_xvalid}
In this appendix we evaluate the predictive performance of four model specifications using a temporal training-test split. Specifically, the training set comprises data from year 1 to 12, while the test set consists of observations from 2022 (\( t = 13 \)), across all regions.
As described in Section \ref{sec:results}, we compare four  Bayesian Beta regression models which differ in two aspects: (i) the specification of the gender effect (fixed vs. random effect), and (ii) the prior distribution assigned to the auto-regressive coefficient $\rho$. The resulting four model variants (named M1, M2, M3 and M4) are summarized in Table \ref{tab:model_variants}, with M1 corresponding to our main model described in Section \ref{sec:methods}.

With respect to gender, the random effect specification models the gender-specific intercept as
\[
\xi_s = \beta_0 + \gamma_s, \quad s = 1, 2,
\]
where $\beta_0$ is the global intercept and $\gamma_s$ represents a random deviation for each gender. The random effects follow a classical hierarchical formulation \citep[e.g.,][]{hoff2009,gelman2006}:
\[
\gamma_s \mid \mu_s, \sigma^2_s \sim \mathcal{N}(\mu_s, \sigma^2_s), \quad
\mu_s \sim \mathcal{N}(0,4), \quad
\sigma^2_s \sim U(1,10).
\]
In contrast, the gender fixed effect specification is the one we describe in Section \ref{subsec:st_model_priors}.

As for the autoregressive coefficient $\rho$, we consider two prior formulations: a $\text{Beta}(1,1)$ prior (constraining $\rho$ to the interval $(0,1)$) as specified in Section \ref{subsec:st_model_priors}, and a standard normal prior $\mathcal{N}(0,1)$, which places no support restrictions and permits values outside the unit interval.

\begin{table}[h!]
\centering
\caption{Summary of model variants.}
\label{tab:model_variants}
\begin{tabular}{@{}lll@{}}
\toprule
\textbf{Model} & \textbf{Gender Effect} & \textbf{Prior on $\rho$} \\
\midrule
M1 & Fixed  & $Beta(1,1)$ \\
M2 & Fixed  & $\mathcal{N}(0,1)$ \\
M3 & Random & $Beta(1,1)$ \\
M4 & Random & $\mathcal{N}(0,1)$ \\
\bottomrule
\end{tabular}
\end{table}

To evaluate the predictive performance, we compute standard error-based metrics for the test set: root mean squared error (RMSE), mean absolute error (MAE), and the sum of the log predictive densities (SLPD). Additionally, we include a measure we refer to as the Bayesian \( p \)-value (BPV), defined for each observation as the proportion of posterior predictive samples exceeding the observed value. Values of BPV near 0.5 indicate that the observation lies near the centre of the predicted distribution, suggesting a good fit. In contrast, values close to 0 or 1 suggest under or over prediction, respectively. We compute a mean BPV (MBPV) across all the test observations as a summary.

\begin{table}[h!]
\centering
\caption{Predictive performance metrics for the four model variants, evaluated on 2022 test data. RMSE (F) and RMSE (M) refer to female and male subgroups, respectively.}
\label{tab:evaluation_metrics}
\begin{tabular}{lrrrrrrr}
\toprule
\textbf{Model} & \textbf{RMSE} & \textbf{RMSE (F)} & \textbf{RMSE (M)} & \textbf{MAE} & \textbf{MBPV} & \textbf{SLPD} \\
\midrule
M1 & 0.010 & 0.010 & 0.010 & 0.008 & 0.537 & 43.72 \\
M2 & 0.013 & 0.012 & 0.014 & 0.010 & 0.717 & 33.17 \\
M3 & 0.010 & 0.009 & 0.010 & 0.008 & 0.571 & 44.62 \\
M4 & 0.014 & 0.013 & 0.015 & 0.012 & 0.757 & 27.78 \\
\bottomrule
\end{tabular}
\end{table}

Table \ref{tab:evaluation_metrics} reports the predictive performance metrics for each model, evaluated on the 2022 test set.
The results suggest that the prior specification for $\rho$ has a greater impact on predictive performance than the treatment of the gender effect. Specifically, models using a Normal prior for $\rho$ (M2, M4) tend to overestimate 2022 obesity rates. This is likely because the unbounded support of the Normal distribution allows $\rho$ to take values greater than one, permitting overly persistent or explosive temporal dynamics and leading to overfitting of upward trends in the training data.
In contrast, the Beta$(1,1)$ prior restricts $\rho$ to the $(0,1)$ interval, constraining the model to more realistic levels of temporal autocorrelation. Models M1 and M3, which use this prior, yield more stable predictions and improved out-of-sample accuracy.
Between M1 and M3, predictive performances are nearly identical. We select M1 for reporting in the main paper, as we think modeling gender as a random effect with only two levels (male and female) is not necessary. A fixed effect specification is more parsimonious in this case.

\begin{figure}[H]
\centering
\begin{subfigure}[b]{0.48\textwidth}
    \centering
    \includegraphics[width=\textwidth]{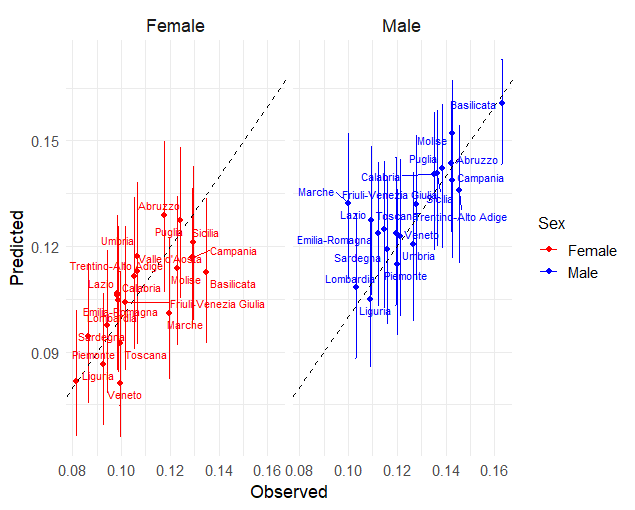}
    \caption{Model M1}
    \label{fig:M1_2022_pred_vs_obs}
\end{subfigure}
\hfill
\begin{subfigure}[b]{0.48\textwidth}
    \centering
    \includegraphics[width=\textwidth]{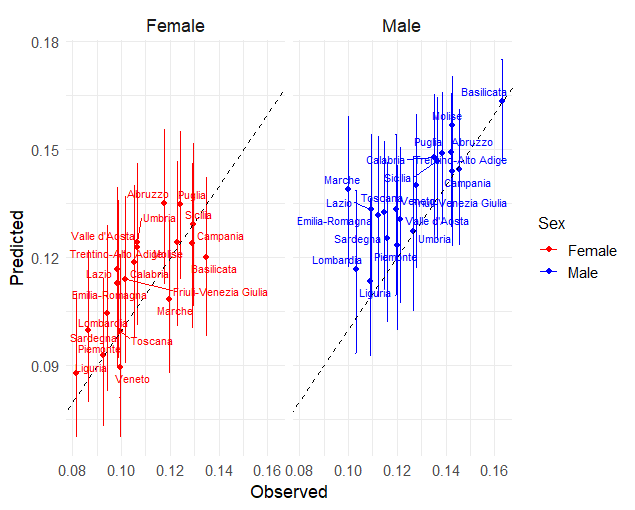}
    \caption{Model M2}
    \label{fig:M2_2022_pred_vs_obs}
\end{subfigure}
\vspace{0.5cm}
\begin{subfigure}[b]{0.48\textwidth}
    \centering
    \includegraphics[width=\textwidth]{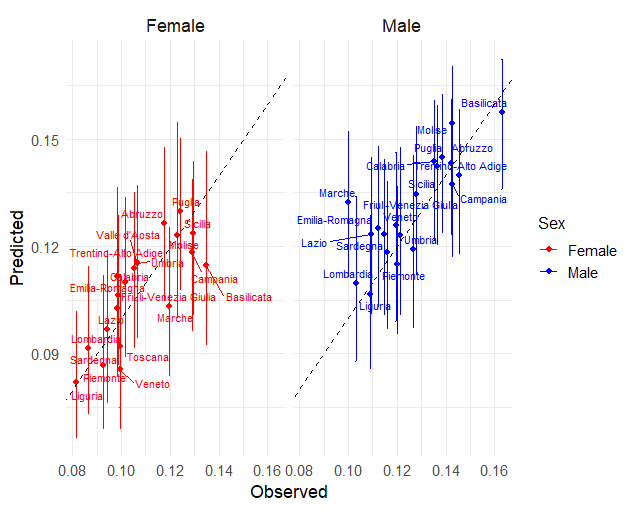}
    \caption{Model M3}
    \label{fig:M3_2022_pred_vs_obs}
\end{subfigure}
\hfill
\begin{subfigure}[b]{0.48\textwidth}
    \centering
    \includegraphics[width=\textwidth]{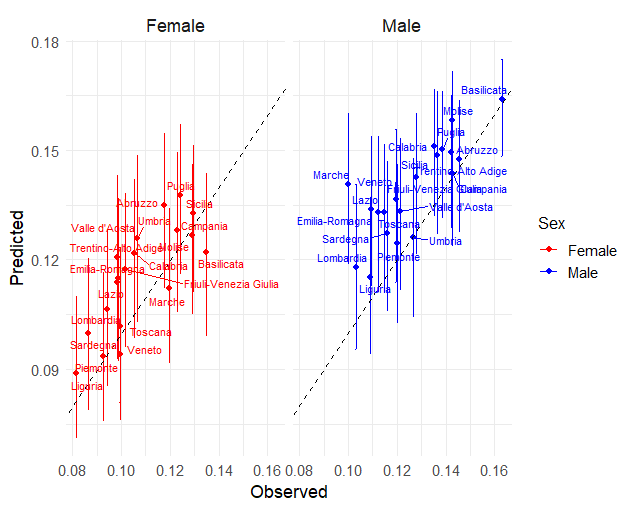}
    \caption{Model M4}
    \label{fig:M4_2022_pred_vs_obs}
\end{subfigure}
\caption{Predicted versus observed obesity rates for 2022 by gender for the four model specifications reported in Table \ref{tab:model_variants}. Points represent posterior means, while vertical bars indicate the 95\% credible intervals.}
\label{fig:2022_pred_vs_obs_all_models}
\end{figure}

In Figures \ref{fig:M1_2022_pred_vs_obs}-\ref{fig:M4_2022_pred_vs_obs} we provide for each of the four models a visual representation of the posterior predictive accuracy by representing the predicted obesity rates against the observed ones, together with 95\% credible intervals, separately by gender.  This helps to assess the uncertainty in the predictions for the four specifications. Overall, Models M2 and M4 display a tendency toward overestimation, particularly for males, as evidenced by the majority of point predictions lying above the 45-degree reference line, together with upward biased intervals. By contrast, M1 and M3 appear more closely aligned with the observed values. In terms of uncertainty, however, the width of the credible intervals is broadly comparable across models, suggesting that differences in performance are mainly driven by bias rather than by substantial changes in predictive variability.

\subsection{Spatial Cross-Validation}\label{subsec:spatial_xvalid}

To further assess the robustness of the proposed modeling strategies and to compare the four model specifications (M1 to M4), we conducted a spatial cross-validation analysis. Unlike the temporal training-test split described in Section \ref{subsec:2022_xvalid}, this validation scheme evaluates the ability of the models to extrapolate to previously unseen geographical areas.
Specifically, we adopt a leave-one-region-out strategy by selecting three representative Italian regions: Lombardia (North), Toscana (Centre), and Campania (South). For each region in turn, all observations corresponding to that region are excluded from the training set, and models M1 to M4 are estimated using data from the remaining regions. Posterior predictive distributions are then generated for the held-out region across all available years.
Predictive accuracy is assessed using the same set of metrics employed in the temporal validation, namely RMSE, MAE, MBPV and SLPD. RMSE is additionally reported separately for female and male subgroups in order to evaluate potential gender-specific discrepancies in predictive performance. 

\begin{table}[h!]
\centering
\caption{Predictive performance metrics by region (all years). RMSE (F) and RMSE (M) refer to female and male subgroups, respectively.}
\label{tab:evaluation_metrics_all_regions}
\begin{tabular}{llrrrrrr}
\toprule
\textbf{Held-out Region} & \textbf{Model} & \textbf{RMSE} & \textbf{RMSE (F)} & \textbf{RMSE (M)} & \textbf{MAE} & \textbf{MBPV} & \textbf{SLPD} \\
\midrule
\multicolumn{8}{l}{\textbf{Lombardia}} \\
\midrule
 & M1 & 0.0086 & 0.0098 & 0.0072 & 0.0070 & 0.654 & 28.8 \\
 & M2 & 0.0086 & 0.0097 & 0.0072 & 0.0068 & 0.651 & 28.7 \\
 & M3 & 0.0089 & 0.0100 & 0.0076 & 0.0074 & 0.660 & 28.3 \\
 & M4 & 0.0089 & 0.0101 & 0.0075 & 0.0074 & 0.660 & 28.3 \\
\midrule
\multicolumn{8}{l}{\textbf{Toscana}} \\
\midrule
 & M1 & 0.0065 & 0.0053 & 0.0075 & 0.0054 & 0.565 & 32.7 \\
 & M2 & 0.0066 & 0.0052 & 0.0077 & 0.0054 & 0.571 & 32.5 \\
 & M3 & 0.0060 & 0.0051 & 0.0068 & 0.0051 & 0.544 & 33.3 \\
 & M4 & 0.0063 & 0.0051 & 0.0072 & 0.0053 & 0.557 & 33.0 \\
\midrule
\multicolumn{8}{l}{\textbf{Campania}} \\
\midrule
 & M1 & 0.0094 & 0.0102 & 0.0086 & 0.0082 & 0.576 & 27.1 \\
 & M2 & 0.0096 & 0.0103 & 0.0089 & 0.0082 & 0.591 & 27.0 \\
 & M3 & 0.0094 & 0.0104 & 0.0083 & 0.0082 & 0.577 & 26.5 \\
 & M4 & 0.0096 & 0.0103 & 0.0090 & 0.0083 & 0.592 & 26.7 \\
\bottomrule
\end{tabular}
\end{table}

Table \ref{tab:evaluation_metrics_all_regions} reports the resulting predictive performance metrics for each model-region combination.
 
Across the three regions, predictive performance is generally higher for Toscana, followed by Lombardia, while Campania exhibits larger prediction errors. This pattern is consistent across model specifications and likely reflects a higher degree of structural heterogeneity or stronger region-specific dynamics in the southern region.
In terms of model comparison, differences between specifications remain modest and do not point to a clear overall winner. While minor variations can be observed, with models M1 and M2 performing slightly better for Lombardia, M3 and M4 for Toscana and M1 slightly outperforming others for Campania, these differences are quantitatively small. This suggests that, when evaluating predictive performance across all years, neither the prior specification for the autoregressive coefficient $\rho$ nor the treatment of the gender effect substantially affects out-of-sample accuracy.
Consequently, modeling $\rho$ using either a Beta or a Normal prior leads to no systematic predictive advantage in this setting, and treating gender as a random effect does not provide clear benefits given the limited number of gender categories. In light of these findings, we favor Model M1, as it offers the most parsimonious specification without sacrificing predictive performance and remains consistent with the results of the temporal 2022 validation discussed in Section \ref{subsec:2022_xvalid}.

\begin{figure}[h]
    \centering
    \includegraphics[width=\textwidth]{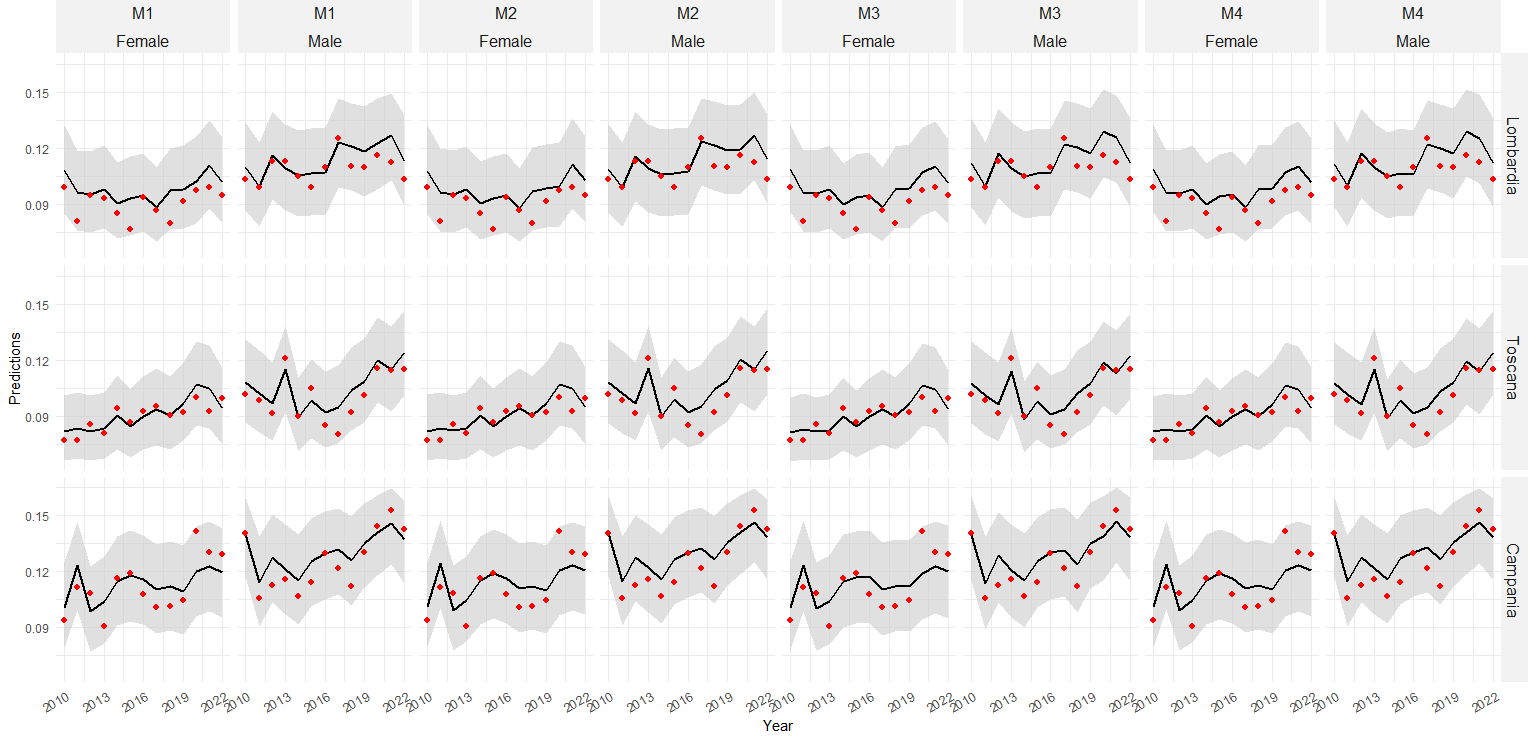}
    \caption{Predicted versus observed obesity rates across all regions (rows) and models (columns), stratified by gender (columns). Shaded areas represent the 95\% credible intervals, the black solid line denotes the posterior mean, and the red points indicate the observed values.}
    \label{fig:allmodels_allreg_pred_vs_obs}
\end{figure}

We present in Figure \ref{fig:allmodels_allreg_pred_vs_obs} the posterior predicted values against the observed data for all held-out regions and models. This visualization makes it possible to assess the uncertainty associated with the predictions across both genders and regions.
Overall, the observed values consistently fall within the corresponding credible intervals. For Toscana (middle row of the plot), the observations are closely aligned with the posterior means, particularly for females, while males exhibit slightly greater predictive uncertainty. For Lombardia (upper row), the observed values are near the predictions but display a minor tendency toward overestimation, in agreement with the MBPV analysis. In Campania (lower row), predictions show higher uncertainty and occasional discrepancies. Notably, for 2020 females across all models, the observed values approach the upper boundary of the 95\% credible interval, suggesting slight underestimation; conversely, other years and genders tend to exhibit mild overestimation, again accordingly to MBPV. In general, predictions remain similar and reliable across regions, and differences between the four model specifications are minimal.

\newpage

\section{Sensitivity Analysis for the SSVS Prior Specification}\label{Appendix4}

We conducted a systematic robustness analysis of the SSVS prior.
We first fixed the slab variance multiplier at $c^2 = 4000$. We then varied the spike variance $\tau^2$ such that the intersection between the spike and slab densities (i.e., $\zeta$ ) took four distinct values: 0.0001, 0.001 (the one we used in the main model (see Section \ref{subsec:ssvs}), 0.01, and 0.1: smaller intersections correspond to tighter spike. The resulting standard deviations for the two components are reported in Table \ref{tab:ssvs_spike_slab_sd}. 

\begin{figure}[ht]
\centering
\caption{Posterior probability evaluation of the inclusion parameters $\omega_k$ for all the covariates according to the four considered values of the intersection parameter $\zeta$.}
\label{fig:ssvs_gamma_post_evaluation_all}

\begin{subfigure}{0.48\textwidth}
    \centering
    \includegraphics[width=\textwidth]{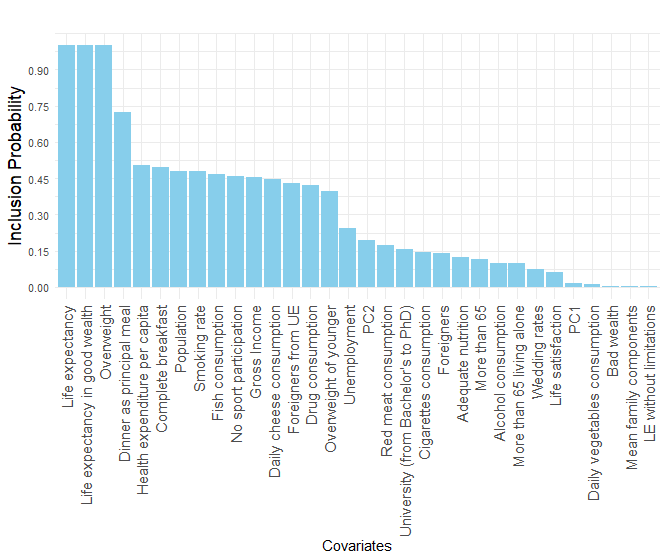}
    \caption{ $\zeta = 0.0001$}
    \label{fig:ssvs_gamma_post_evaluation_0.0001}
\end{subfigure}
\hfill
\begin{subfigure}{0.48\textwidth}
    \centering
    \includegraphics[width=\textwidth]{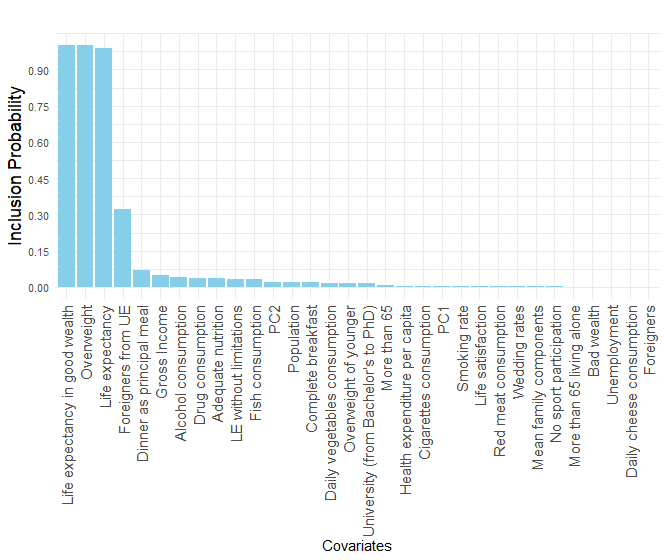}
    \caption{$\zeta = 0.001$}
    \label{fig:ssvs_gamma_post_evaluation_0.001}
\end{subfigure}

\vspace{0.4cm}

\begin{subfigure}{0.48\textwidth}
    \centering
    \includegraphics[width=\textwidth]{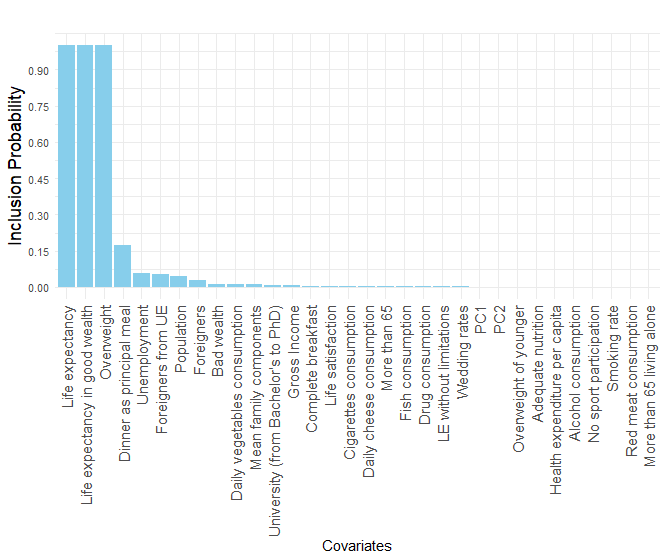}
    \caption{$\zeta= 0.01$}
    \label{fig:ssvs_gamma_post_evaluation_0.01}
\end{subfigure}
\hfill
\begin{subfigure}{0.48\textwidth}
    \centering
    \includegraphics[width=\textwidth]{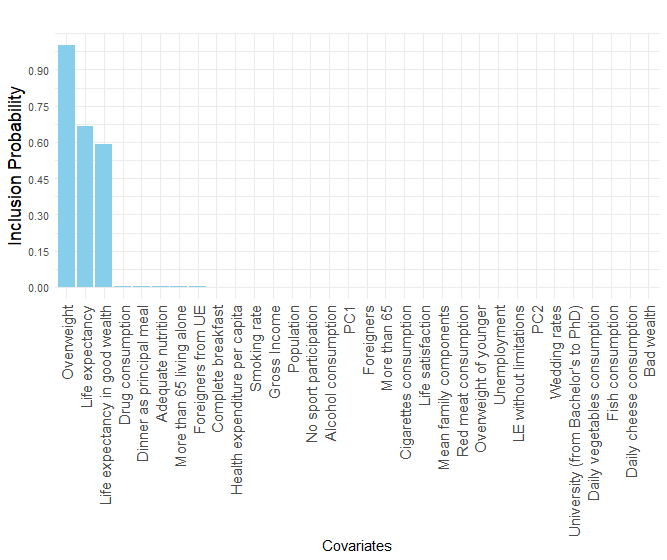}
    \caption{$\zeta= 0.1$}
    \label{fig:ssvs_gamma_post_evaluation_0.1}
\end{subfigure}

\end{figure}

For each specification, we re-estimated the model presented in Section \ref{subsec:st_model_likelihood} (denoted as M1). 
For what concerns the SSVS indicators $\omega_k$ and their associated inclusion probabilities $\theta_k$, we set all initial values equal to 1, i.e. starting the search from the full model. We let the sampler run for 20,000 iterations, with a burn-in of 5,000 and no thinning.
We then computed the posterior inclusion probabilities $\omega_k$ for each covariate, which are displayed in Figures \ref{fig:ssvs_gamma_post_evaluation_0.0001}-\ref{fig:ssvs_gamma_post_evaluation_0.1}.

Across all prior configurations, the same three covariates consistently exhibit the highest posterior inclusion probabilities: life expectancy, life expectancy in good health and overweight prevalence. This concordance across substantially different shrinkage regimes indicates that these covariates capture robust empirical signal rather than reflecting a particular prior choice. Nonetheless, the absolute magnitude of inclusion probabilities varies with the strength of the prior separation. Under the strongest spike concentration ($\zeta = 0.0001$), overall inclusion probabilities increase, and several additional covariates attain intermediate probability levels (still below 0.5), with the only exception of the covariate \enquote{dinner as principal meal} reaching a posterior inclusion probability slightly above 0.7. The two intermediate separation settings ($\zeta$ equal to 0.001 and 0.01) produce almost identical results, both characterised by substantial shrinkage of most coefficients. The weakest separation ($\zeta= 0.1$) yields the lowest inclusion probabilities across all covariates: the same three covariates identified in the main analysis remain the most influential, while all others are effectively excluded, in line with the findings of \citet{rockova2012}.

These results suggest that although the range of the inclusion probabilities is somehow affected by the strength of prior shrinkage, the variable ranking is highly robust with respect to the SSVS hyperparameter choice.

\begin{table}[h!]
\centering
\caption{Spike and slab standard deviations for the four SSVS prior specifications according to the intersection value $\zeta$.}
\label{tab:ssvs_spike_slab_sd}
\begin{tabular}{lcccc}
\toprule
\textbf{Intersection} $\zeta$  & \textbf{0.0001} & \textbf{0.001} & \textbf{0.01} & \textbf{0.1} \\
\midrule
\textbf{Spike SD} & 0.00002 & 0.00024 & 0.00245 & 0.02455 \\
\textbf{Slab SD} & 0.098 & 0.982 & 9.821 & 98.21 \\
\bottomrule
\end{tabular}
\end{table}

\FloatBarrier

\section{Model Diagnostics}\label{Appendix5}
For the model presented in Section \ref{subsec:st_model_likelihood}, 
we conducted a comprehensive set of diagnostic checks to assess MCMC convergence, sampling efficiency and model adequacy. First, we evaluated convergence and effective sample size using the diagnostic frameworks proposed by \citet{coda2006} and \citet{vehtari2021}. Second, we examined residual patterns to detect potential model misspecification. Finally, we performed posterior predictive checks to assess the model’s ability to reproduce key features of the observed data.

\subsection{Convergence Diagnostics}\label{subsec:conv_diagn}

We assessed MCMC convergence and sampling efficiency using both classical and modern diagnostic tools.
Specifically, we report the traditional potential scale reduction factor $\widehat{R}$ and the associated effective sample size (ESS) as implemented in the \texttt{coda} package \citep{coda2006}, based on the original proposals of \citet{gelman_rubin_1992} and \citet{gelman_brooks_1998}. 
We also use the rank-normalized split-$\widehat{R}$ together with the basic, bulk, and tail ESS introduced by \citet{vehtari2021} and implemented in the \texttt{posterior} package.

We report the resulting convergence diagnostics in Table \ref{tab:ess_rhat} for the intercept $\beta_0$ and the male-specific intercept component $\gamma$, the spatial and temporal random effects $\bm{\psi}$ and $\bm{\alpha}$, the autoregressive coefficient $\rho$, the precision parameter $\phi$ and the random effect precision hyperparameters $\tau_{\psi}$ and $\tau_{\alpha}$.
SSVS parameters and fixed effect coefficients (i.e., $\bm \beta$) are excluded from the analysis because their SSVS hierarchical specification makes more difficult the analysis of the two chains.

Overall, effective sample sizes are generally large under both classical and more conservative diagnostics \citep{vehtari2021}, with only a few spatial and late-period temporal random effects showing moderate reductions. Convergence diagnostics based on $\widehat{R}$ (including rank-normalized split-$\widehat{R}$) are uniformly close to one, providing no evidence of lack of convergence; taken together, these results indicate satisfactory mixing and support reliable posterior inference.

\mycomment{Overall, the classical ESS values computed via \texttt{coda} are consistently well above 100 for all parameters and often substantially larger, suggesting adequate sampling efficiency when assessed using traditional criteria.
In contrast, the basic, bulk, and tail ESS proposed by \citet{vehtari2021} are noticeably more conservative but still greater than 100 for almost all the parameters.
Only for a small subset of parameters, specifically a few spatial random effects (e.g., $\psi_2$ and $\psi_{15}$) and several of the temporal random effects for the last years, the bulk or tail ESS falls below 100, indicating reduced efficiency in exploring either the center or the tails of the posterior distribution.
Taken together, these results indicate that while mixing is not uniformly efficient across all the parameters, it  is still acceptable for reliable posterior inference, as we note also in Section \ref{subsec:bayes_resid_analys} and \ref{subsec:ppc}).
Convergence diagnostics based on $\widehat{R}$ and on the rank-normalized split-$\widehat{R}$ values are equal to 1.00 or 1.01 for nearly all parameters, reaching 1.03 only in a few cases regarding the spatial and temporal random effects.
Similarly, the classical $\widehat{R}$ values computed via $\texttt{coda}$ are all close to unity, with 95\% upper confidence bounds remaining well below alarming thresholds. These results provide evidence against lack of convergence.
Focusing on specific parameter blocks, the intercept parameters $\beta_0$ and $\gamma$ exhibit acceptable mixing, with bulk and tail ESS well above 100 and split-$\widehat{R}$ equal to 1.01.
The spatial random effects $\psi_1,\ldots,\psi_{20}$ show stable convergence behavior, with $\widehat{R}$ values essentially equal to one and generally large ESS, despite mild inefficiencies for a small number of components.
The temporal random effects $\alpha_1,\ldots,\alpha_{13}$ also converge satisfactorily, although their ESS values are comparatively smaller, reflecting stronger autocorrelation.
The autoregressive coefficient $\rho$ mixes very well, the precision parameter $\phi$ and the spatial precision hyperparameters $\tau_{\psi}$ exhibit excellent convergence and very large effective sample sizes, while the temporal precision hyperparameter $\tau_{\alpha}$ shows lower but acceptable ESS's.
Although ESS diagnostics by \citet{vehtari2021} reveal some localized inefficiencies, the overall diagnostics support convergence to the target posterior distribution.
}

\begin{table}[h!]
\centering
\caption{Effective sample size (ESS) and convergence diagnostics for model parameters. 
ESS (basic), ESS (bulk), ESS (tail), and rank-normalized split (rn-split) $\widehat{R}$ are computed following \citet{vehtari2021} using the \texttt{posterior} package. 
Classical $\widehat{R}$, its 95\% upper confidence bound (UCB), and ESS are computed using the \texttt{coda} package \citep{coda2006}. 
}
\label{tab:ess_rhat}
\begin{tabular}{lrrrrrrr}
\hline
Parameter 
& ESS (basic) 
& ESS (bulk) 
& ESS (tail) 
& rn-split $\widehat{R}$ 
& ESS 
& $\widehat{R}$ 
& 95\% UCB $\widehat{R}$ \\
\hline
$\beta_0$   & 110   & 112   & 189   & 1.01 & 122   & 1.01 & 1.02 \\
$\gamma$    & 169   & 169   & 424   & 1.01 & 167   & 1.01 & 1.04 \\

$\psi_{1}$  & 121  & 125  & 242   & 1.01 &  813 & 1.01 & 1.03 \\
$\psi_{2}$  & 76   & 95   & 49    & 1.01 &  286 & 1.03 & 1.07 \\
$\psi_{3}$  & 126  & 126  & 804   & 1.00 &  273 & 1.01 & 1.03 \\
$\psi_{4}$  & 284  & 289  & 975   & 1.00 & 1120 & 1.00 & 1.00 \\
$\psi_{5}$  & 151  & 152  & 536   & 1.00 &  361 & 1.01 & 1.01 \\
$\psi_{6}$  & 190  & 192  & 330   & 1.00 & 1068 & 1.00 & 1.00 \\
$\psi_{7}$  & 386  & 385  & 1761  & 1.00 &  780 & 1.00 & 1.01 \\
$\psi_{8}$  & 142  & 143  & 528   & 1.00 &  280 & 1.00 & 1.00 \\
$\psi_{9}$  & 548  & 549  & 3185  & 1.00 & 1002 & 1.00 & 1.00 \\
$\psi_{10}$ & 817  & 819  & 3765  & 1.01 & 1425 & 1.00 & 1.02 \\
$\psi_{11}$ & 408  & 408  & 1336  & 1.00 &  553 & 1.00 & 1.01 \\
$\psi_{12}$ & 342  & 342  & 4145  & 1.01 &  802 & 1.00 & 1.00 \\
$\psi_{13}$ & 1007 & 1014 & 10691 & 1.00 & 4025 & 1.00 & 1.00 \\
$\psi_{14}$ & 187  & 190  & 383   & 1.00 &  721 & 1.01 & 1.01 \\
$\psi_{15}$ & 51   & 56   & 67    & 1.03 &  203 & 1.03 & 1.09 \\
$\psi_{16}$ & 90   & 93   & 98    & 1.01 &  597 & 1.01 & 1.01 \\
$\psi_{17}$ & 162  & 162  & 508   & 1.00 &  595 & 1.00 & 1.00 \\
$\psi_{18}$ & 129  & 131  & 284   & 1.01 &  622 & 1.01 & 1.01 \\
$\psi_{19}$ & 74   & 77   & 165   & 1.01 &  257 & 1.01 & 1.02 \\
$\psi_{20}$ & 208  & 212  & 487   & 1.01 & 1444 & 1.02 & 1.10 \\

$\alpha_{1}$  & 115 & 121 & 236 & 1.02 & 193 & 1.00 & 1.00 \\
$\alpha_{2}$  & 130 & 135 & 215 & 1.02 & 200 & 1.01 & 1.01 \\
$\alpha_{3}$  & 101 & 104 & 204 & 1.02 & 166 & 1.02 & 1.07 \\
$\alpha_{4}$  & 81  & 83  & 212 & 1.02 & 181 & 1.03 & 1.11 \\
$\alpha_{5}$  & 82  & 83  & 221 & 1.02 & 178 & 1.03 & 1.13 \\
$\alpha_{6}$  & 139 & 142 & 239 & 1.01 & 222 & 1.01 & 1.02 \\
$\alpha_{7}$  & 74  & 75  & 216 & 1.03 & 199 & 1.03 & 1.14 \\
$\alpha_{8}$  & 73  & 75  & 206 & 1.03 & 190 & 1.03 & 1.14 \\
$\alpha_{9}$  & 64  & 66  & 216 & 1.03 & 185 & 1.04 & 1.15 \\
$\alpha_{10}$ & 62  & 63  & 226 & 1.03 & 188 & 1.03 & 1.15 \\
$\alpha_{11}$ & 67  & 66  & 221 & 1.03 & 186 & 1.03 & 1.11 \\
$\alpha_{12}$ & 72  & 72  & 232 & 1.02 & 196 & 1.02 & 1.10 \\
$\alpha_{13}$ & 83  & 84  & 213 & 1.02 & 181 & 1.02 & 1.09 \\

$\rho$          & 506   & 484   & 2784  & 1.01 & 720   & 1.00 & 1.01 \\
$\phi$          & 10338 & 10320 & 11275 & 1.00 & 10255 & 1.00 & 1.00 \\
$\tau_{\psi}$   & 9396  & 9298  & 11238 & 1.00 & 10454 & 1.00 & 1.00 \\
$\tau_{\alpha}$ & 176   & 187   & 1079  & 1.01 & 617   & 1.01 & 1.06 \\
\hline
\end{tabular}

\end{table}

\FloatBarrier

\subsection{Residual Analysis}\label{subsec:bayes_resid_analys}

We conducted a comprehensive residual analysis to assess the adequacy of the model and to identify potential systematic patterns not captured by the specified model structure. We define residuals as the difference between the observed outcomes and the posterior mean of the predictive distribution (as given by M1).

\begin{figure}[h!]
    \centering
    \includegraphics[scale=0.5]{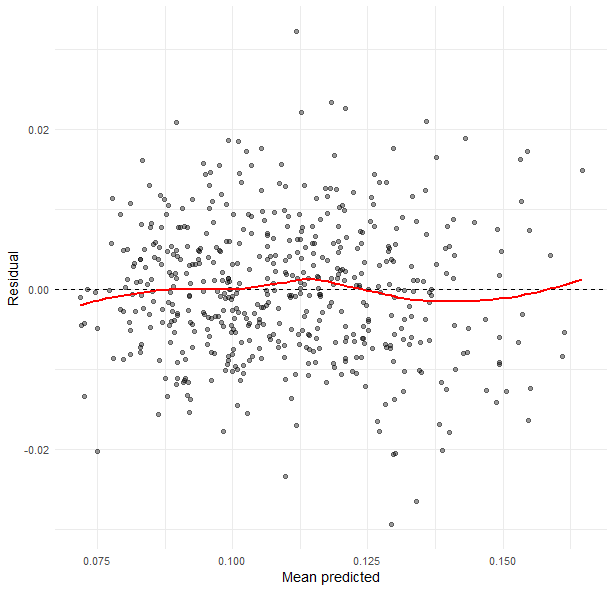}
    \caption{Residuals versus obesity rate predictions (posterior mean). The solid curve represents a smooth fitted function.}
    \label{fig:resid_vs_pred}
\end{figure}

Figure \ref{fig:resid_vs_pred} displays the residuals as a function of the posterior mean predictions. The residual cloud is symmetrically distributed around zero, with no discernible structure or systematic trend across the range of predicted values. The smooth fitted curve remains close to zero, indicating the absence of systematic structure in the residuals. 

\begin{figure}[h!]
    \centering
    \includegraphics[width=\textwidth]{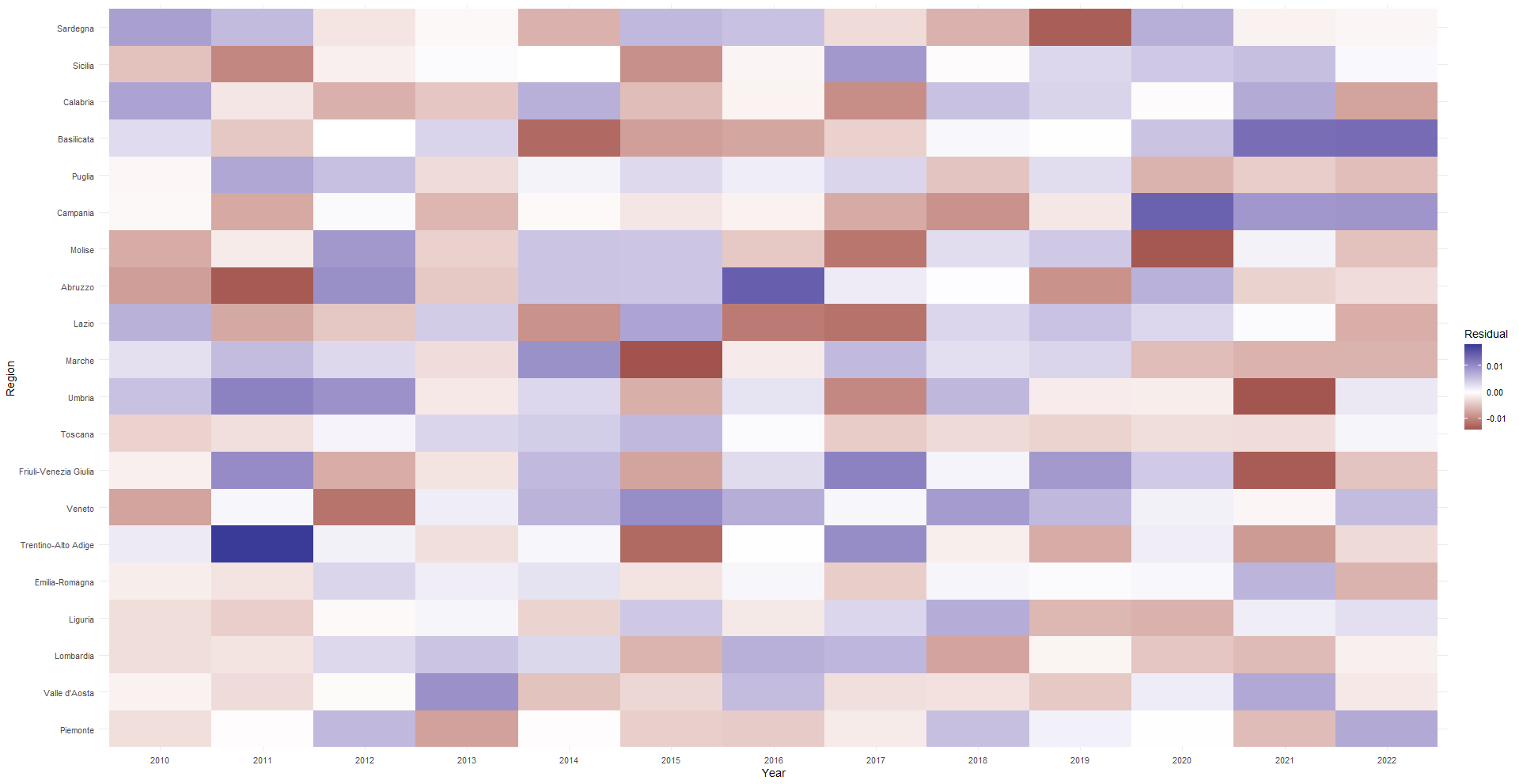}
    \caption{Heatmap of mean posterior residuals by region and year. Each cell represents the average posterior mean residual for a given region-year combination. The color scale is centered at zero, with warmer (cooler) colors indicating negative (positive) residuals.}
    \label{fig:heatmap}
\end{figure}

We additionally summarize residual behavior jointly across space and time using a heatmap representation. Figure \ref{fig:heatmap} displays the mean posterior residual for each region-year combination, obtained by averaging posterior mean residuals within each spatial unit and year. The color scale is centered at zero, such that positive (negative) values indicate systematic underprediction (overprediction) by the model.
The heatmap provides a compact diagnostic for detecting localized or persistent model misspecification across regions or over time. Overall, residuals appear small in magnitude and are distributed symmetrically around zero across both dimensions. No region exhibits sustained positive or negative residuals over consecutive years, nor is there evidence of a coherent temporal trend shared across regions. Instead, residual deviations appear sporadic and weakly structured, suggesting that the model adequately captures the dominant spatio-temporal patterns in the data.

\mycomment{Finally, to assess whether the model adequately captured temporal dependence, we conducted a Bayesian residual autocorrelation analysis based on the posterior predictive distribution. For each spatial unit, we computed the residuals at every time point as the difference between the observed outcome and each posterior predictive draw. We then calculated the autocorrelation function (ACF) for each posterior draw and spatial unit, yielding a posterior distribution of the ACF at each lag.
For each space-lag combination, we summarized this distribution by its posterior mean and corresponding 95\% credible interval. Figure \ref{fig:resid_acf} displays these summaries, with solid lines representing the mean ACF and shaded bands denoting the 95\% credible intervals, faceted by spatial unit. 
For all regions, the posterior credible intervals of the residual ACF include zero at all lags, indicating no evidence of remaining temporal autocorrelation. This suggests that the temporal dynamics specified in the model is generally sufficient to account for the serial dependence in the data.

\begin{figure}[h!]
    \centering
    \includegraphics[width=\textwidth]{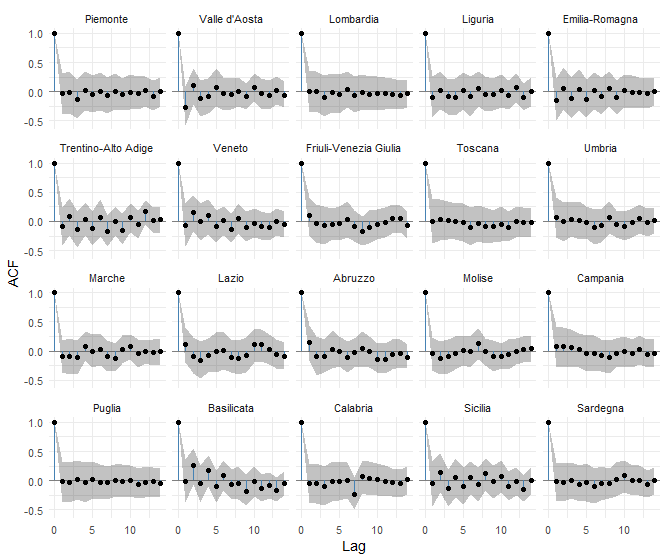}
\caption{Posterior summaries of the residual autocorrelation function (ACF) by spatial unit. Black dots show the posterior mean ACF, while shaded regions denote 95\% credible intervals computed from posterior predictive residuals. The horizontal grey line indicates zero autocorrelation.}
\label{fig:resid_acf}
\end{figure}
}

\FloatBarrier

\subsection{Posterior Predictive Checks}\label{subsec:ppc}

\begin{figure}[h!]
    \centering
    \includegraphics[width=\textwidth]{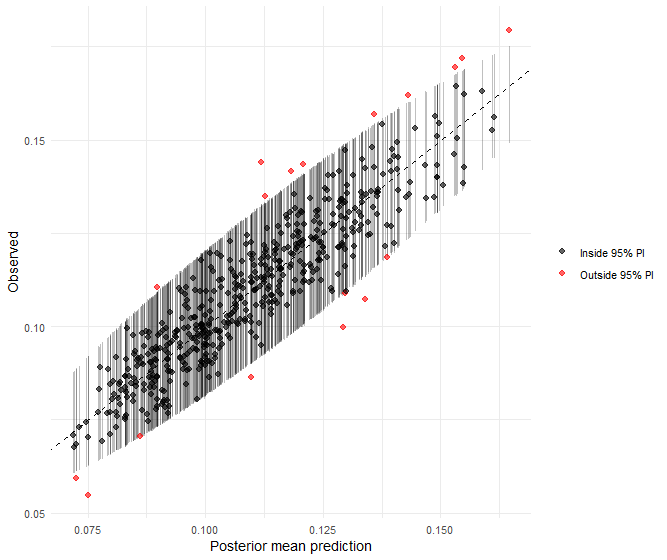}
    \caption{Posterior predictive check comparing observed outcomes to posterior mean predictions. Points represent individual observations, vertical bars denote 95\% posterior predictive intervals, and the dashed line indicates the identity line. Red points denote the prediction means falling outside the 95\% credible interval.}
    \label{fig:ppc}
\end{figure}

We further evaluated model adequacy using posterior predictive checks based on the full posterior predictive distribution. For each observation, we summarized the posterior predictive draws by computing the posterior predictive mean and the 2.5\%, 50\%, and 97.5\% predictive quantiles. These summaries were then compared with the corresponding observed outcomes.
Figure \ref{fig:ppc} displays the observed values against the posterior predictive mean predictions, with vertical error bars representing the 95\% posterior predictive intervals. The dashed diagonal line indicates perfect agreement between predictions and observations. Overall, the observed data align closely with the posterior predictive means, with points symmetrically scattered around the identity line and no visible systematic deviations across the range of predicted values. The width of the predictive intervals appears relatively stable across observations.
Quantitatively, the vast majority of observations fall within their corresponding 95\% posterior predictive intervals, indicating good overall predictive calibration. Only a small subset of observations lies outside these intervals: specifically, 18 out of 520 observations (approximately 3.5\%) fall below the 2.5\% or above the 97.5\% posterior predictive quantiles.
Overall, the limited number, moderate magnitude, and heterogeneous distribution of these exceedances support the conclusion that the model adequately captures both the central tendency and the uncertainty of the observed outcomes.

\mycomment{

\begin{table}[h!]
\centering
\caption{Observations falling outside the 95\% posterior predictive intervals.}
\label{tab:ppc_outliers}
\begin{tabular}{llllll}
\hline
Observed & Predicted mean & Residual & Sex & Region & Year \\
\hline
0.0592 & 0.0726 & $-0.0134$ & Female & Liguria & 2011 \\
0.0548 & 0.0750 & $-0.0202$ & Female & Veneto & 2012 \\
0.1100 & 0.0897 &  0.0208  & Female & Veneto & 2018 \\
0.1420 & 0.1180 &  0.0233  & Female & Campania & 2020 \\
0.1570 & 0.1360 &  0.0210  & Female & Basilicata & 2021 \\
0.1350 & 0.1130 &  0.0221  & Female & Basilicata & 2022 \\
0.0863 & 0.1100 & $-0.0234$ & Female & Sicilia & 2011 \\
0.1440 & 0.1120 &  0.0322  & Male & Trentino-Alto Adige & 2011 \\
0.1430 & 0.1210 &  0.0226  & Male & Trentino-Alto Adige & 2017 \\
0.0705 & 0.0862 & $-0.0157$ & Male & Veneto & 2010 \\
0.1190 & 0.1390 & $-0.0201$ & Male & Friuli-Venezia Giulia & 2021 \\
0.1090 & 0.1300 & $-0.0206$ & Male & Marche & 2015 \\
0.0999 & 0.1290 & $-0.0294$ & Male & Marche & 2022 \\
0.1690 & 0.1530 &  0.0162  & Male & Abruzzo & 2020 \\
0.1620 & 0.1430 &  0.0188  & Male & Molise & 2012 \\
0.1800 & 0.1650 &  0.0149  & Male & Molise & 2021 \\
0.1070 & 0.1340 & $-0.0265$ & Male & Basilicata & 2014 \\
0.1720 & 0.1550 &  0.0172  & Male & Sicilia & 2021 \\
\hline
\end{tabular}
\end{table}

}

\end{document}